\documentclass[]{jfm}

\usepackage{amsmath}
\usepackage{psfig}
\usepackage{fancyhdr}
\usepackage{rotate}
\usepackage{rotating}
\usepackage{dcolumn}
\usepackage{bm}
\usepackage{subfigure}
\usepackage{epsfig}
\usepackage{natbib}
\usepackage{graphicx}
\usepackage{color}
\usepackage{latexsym}
\usepackage{amsmath}

\newcommand{\vect}[1]{\mbox{\boldmath$#1$\unboldmath}}

\newcommand{\M}{mol~l$^{-1}$}

\newcommand{\degc}{$^\circ$C}

\newcommand{\ie}{{\it i.e.}}
\newcommand{\eg}{{\it e.g.}}

\newcommand{\bsl}{{\ell}}

\newcommand{\Pe}{\mbox{Pe}}

\newcommand{\kb}{k}
\newcommand{\ka}{\kappa a}

\newcommand{\grad}{\vect{\nabla}}
\newcommand{\lapl}{\nabla^2}
\newcommand{\dive}{\vect{\nabla} \cdot}

\title[Transport in polymer-gel composites]{Transport in polymer-gel
composites: Theoretical methodology and response to an electric field}

\author[]{R\ls E\ls G\ls H\ls A\ls N\ns J.\ns H\ls I\ls L\ls L}

\affiliation{Department of Chemical Engineering and McGill Institute
  for Advanced Materials, McGill University, Montreal, Quebec, H3A2B2,
  CANADA}

\begin{document}

\maketitle

\begin{abstract}

A theoretical model of electromigrative, diffusive and convective
transport polymer-gel composites is presented. Bulk properties are
derived from the standard electrokinetic model with an impenetrable
charged sphere embedded in an electrolyte-saturated Brinkman
medium. Because the microstructure can be carefully controlled, these
materials are promising candidates for enhanced gel-electrophoresis,
chemical sensing, drug delivery, and microfluidic pumping
technologies. The methodology provides solutions for situations where
perturbations from equilibrium are induced by gradients of
electrostatic potential, concentration and pressure. While the volume
fraction of the inclusions should be small, Maxwell's well-known
theory of conduction suggests that the model may also be accurate at
moderate volume fractions. In this work, the theory is used to compute
ion fluxes, electrical current density, and convective flow driven by
an electric field applied to an homogeneous composite. The
electric-field-induced (electroosmotic) flow is a sensitive indicator
of the inclusion $\zeta$-potential and size, electrolyte
concentration, and Darcy permeability of the gel, while the electrical
conductivity is most often independent of the polymer gel and is
relatively insensitive to characteristics of the inclusions and
electrolyte.

\end{abstract}


\section{Introduction} \label{sec:introduction}

Gel-electrophoresis is widely used to sort macromolecules based on
their size and electrical charge. Selectivity to size is achieved by
adjusting the permeability of the gel (\eg, agarose or polyacrylamide)
through the concentrations of monomer, cross-linker, catalyst and
initiator used in the gel synthesis. Molecular sorting based on other
characteristics, such as receptor-ligand binding affinity, requires
the gel to exhibit specific physicochemical activity. One way to
achieve this in a controlled manner is to embed surface-functionalized
particles (\eg, biological cells, and synthetic polymer or silica
spheres) in a conventional polymer gel.  Accordingly, this work seeks
to quantify the influence of surface charge on transport driven by
gradients of chemical and electrostatic potential. While the task is
simplified to some extent by limiting the analysis to simple
electrolytes whose mobilities are unhindered by the polymer gel, the
methodology provides a significant step toward a theory that also
accounts for hindered transport of larger electrolyte ions (\eg,
proteins and DNA fragments).

This work also provides a quantitative interpretation of novel
diagnostic tests---analogous to well-established microelectrophoresis
and conductivity measurements---that probe the surface charge or
$\zeta$-potential of immobilized colloids in electrolytes where the
particles would otherwise aggregate. Attractive particle-interaction
potentials arise when the solution pH approaches the isoelectric point
of the particle-electrolyte interface, or the surface charge is
sufficiently well screened by added salt~\citep{Russel:1989}. By
immobilizing colloids in an (ideally) inert (uncharged) polymer gel,
at a pH and ionic strength where the interactions are repulsive, the
pH and ionic strength may be varied without inducing coagulation.

Membranes of sintered glass beads (without intervening polymer) have
long been used in ion-selective electrodes, and, more recently, as
electroosmotic pumps~\citep[\eg,][]{Yao:2003b}. Their simple design
(no moving parts) and high-pressure low-flow characteristics are
ideally suited to microfluidic applications.  Filling the void space
with a permeable, uncharged polymer gel, as proposed in this work,
will increase viscous dissipation and, therefore, diminish pumping
efficiency. Nevertheless, because applications are envisioned where
poor pumping efficiency might be tolerated in view of other
attributes, a quantitative analysis of electroosmotic pumping is
undertaken here.

The charge on particles dispersed in an electrolyte endows them with
an electrophoretic mobility~\citep{Russel:1989}. Theoretical
interpretation of the mobility~\citep{OBrien:1978}, sedimentation
potential~\citep{Saville:1982}, low-frequency
conductivity~\citep{Saville:1979,OBrien:1981}, dielectric response
(complex conductivity)~\citep{DeLacey:1981}, and electroacoustic
response (dynamic mobility)~\citep{OBrien:1988,OBrien:1990} is widely
used to infer the surface charge and, therefore, to indicate
dispersion stability. Closely related are streaming-potential and
streaming-current devices, which are used to infer the charge on
macroscopic substrates, and the charge density and permeability of
porous plugs and coatings~\citep[\eg,][]{Hunter:2001,Dukhin:2004}.

This paper addresses a new but related problem in which impenetrable
spheres with surface charge are randomly dispersed and immobilized in
a permeable, electrolyte-saturated polymer gel (see
figure~\ref{fig:figure1}). The transport processes that take place
with the application of average (macroscale) gradients of
electrostatic potential, electrolyte concentration and pressure are
derived.  While transport of electrolyte ions is relatively
straightforward to calculate in an homogeneous, uncharged gel, charged
inclusions disturb the applied fields, and applied fields disturb the
equilibrium state of the diffuse double layers, so the resulting
fluxes reflect a complex coupling of electromigration, diffusion, and
convective transport.

Electrokinetic theories are often based on the {\em standard
electrokinetic model}~\citep{Overbeek:1943,Booth:1950}, whereby
continuum equations governing the electric field, mobile charge
(microions), mass, and momentum are solved with appropriate boundary
conditions. The principal difficulty usually lies in capturing
double-layer polarization and relaxation at surfaces whose radius of
curvature is comparable to or smaller than the equilibrium
double-layer thickness (Debye length). For `bare' and `soft'
(polymer-coated) particles, polarization and relaxation can be
addressed with novel numerical
methodologies~\citep{OBrien:1978,Hill:2003a}. In this work, a
numerically exact solution of the problem is achieved for the
low-volume-fraction limit where particle interactions can be
neglected. The methodology also neglects quadratic and higher-order
perturbations to the equilibrium state, providing {\em asymptotic
coefficients} that characterize the far-field (power-law) decays of
the velocity disturbance and perturbations to the equilibrium electric
field and ion concentrations. In turn, the asymptotic coefficients are
linked to bulk properties of the composite. The analysis resembles
Maxwell's theory for the effective conductivity of a dilute, random
configuration of spherical inclusions.

{\em Dipolar disturbances}\footnote{Here, {\em dipolar} refers to
axisymmetric disturbances, without a source, that satisfy Laplace's
equation; these take the form $\alpha_j \partial r^{-1} / \partial x_j
= -\alpha_j x_j r^{-3}$, where $r = |\vect{x}|$ and $\alpha_i$ are the
components of a constant vector.} arise in the limit where the
inclusions are uncharged. Then, with the application of a uniform
electric field, the perturbed electrostatic potential $\psi'$ reflects
the non-conducting (impenetrable) surface of the inclusions,
satisfying $\lapl{\psi'} = 0$. Similarly, when subjected to a bulk
concentration gradient, the perturbed ion concentrations $n'$ satisfy
$\lapl{n'} = 0$. In both cases, there is no convective flux, because
the electrolyte is everywhere electrically neutral. With a uniform
pressure gradient $\langle \grad p \rangle$, the problem simplifies to
the flow of an incompressible Newtonian fluid through a Brinkman
medium with impenetrable inclusions. The (solenoidal) velocity
$\vect{u}$ satisfies $\eta \lapl{\vect{u}} - \grad{p} - (\eta /
\bsl^2) \vect{u} = \vect{0}$, where $p$ is the pressure, $\eta$ is the
fluid viscosity, and $\bsl$ is the Brinkman screening length (square
root of the Darcy permeability). When the inclusion radius $a \gg
\bsl$, $\vect{u} = -(\bsl^2 / \eta) \grad{p}$ with $\lapl{p} = 0$
(Darcy flow), and the drag force on the inclusions is $2 \pi \eta
\vect{U} a^3 \bsl^{-2}$, where $\vect{U} = - (\bsl^2 / \eta) \langle
\grad p \rangle + O(\phi)$ is the average fluid velocity. In this
limit, the inclusion contribution to the average drag force (per unit
volume) of the {\em composite} is $n 2 \pi \eta \vect{U} a^3 \bsl^{-2}
= (3/2) \phi \eta \vect{U} \bsl^{-2}$, where $n$ is the inclusion
number density and $\phi = n (4/3) \pi a^3$ is the volume
fraction. With the same far-field velocity, the drag force on each
inclusion is clearly much greater than the Stokes drag force $6 \pi
\eta a \vect{U}$. Note also that the velocity disturbances $\vect{u}'
= \vect{u} - \vect{U} = -(1/2) a^3 U_j (\delta_{ij} r^{-3} - 3 x_i x_j
r^{-5})$ decay as $r^{-3}$ when the distance from the inclusions $r
\gg a \gg \bsl$.

When the inclusions are charged, a diffuse layer of mobile
counter-ions envelops each inclusion, and electroneutrality demands
that the net charge in the diffuse layers balances the immobile
surface charge. With an applied electric field, the electrical body
force within the diffuse layers drives an `inner' flow which, in turn,
drives an $O(\phi)$ `outer' flow, with an $O(1)$ contribution due to
an imposed pressure gradient.  The electric-field-induced velocity
disturbances `pump' fluid through the polymer gel without exerting a
net force on the composite. While the velocities are very low,
composites with an a 1~cm$^2$ cross-section can produce velocities of
several microns per second in a microchannel. Note that a pressure
gradient is necessary to overcome the drag required to pump fluid
through an external network. However, the pressure-driven contribution
to the flow is often small compared to the electric-field-induced
flow. Under these conditions, the electric-field-induced flow rate is
practically independent of the pressure gradient, and the maximum
pressure gradient that can be sustained is limited by the strength of
the composite and its support.

Ion fluxes manifest in an electrical current and, hence, a measurable
{\em electrical conductivity}. Following earlier treatments of the
low-frequency conductivity of dilute colloidal
dispersions~\citep{Saville:1979,OBrien:1981}, the incremental
contribution of the inclusions to the effective conductivity is
calculated. These results link conductivity measurements to the
particle surface charge density, for example. Because the fluxes are
dominated by electromigration, the conductivity is not significantly
influenced by the gel.

The model also provides the effective diffusivities of electrolyte
ions when the bulk electrolyte concentration varies slowly in space
and time.  In a forthcoming paper (to be published elsewhere), two
important situations are examined: ($i$) bulk diffusion in the absence
of an average electric field, and ($ii$) bulk diffusion with an
electric field yielding zero current density. The former provides a
simple setting in which to demonstrate the influence of the inclusions
on the effective ion diffusion coefficients, whereas the latter
provides the particle contribution to the {\em membrane diffusion
potential}, which is well known in the fields of membrane biology,
electrochemistry, and electrochemical engineering. In both cases, the
particle contribution to these bulk properties may be comparable to or
larger than in the absence of inclusions.

The paper is organized as follows. We begin in~\S\ref{sec:theory} with
a description of the electrokinetic model. First, the composite
microstructure and an apparatus for comparing theory and experiment
are described. The subsections therein present the electrokinetic
transport equations (\S\ref{sec:equations}), which are used to compute
linearly independent solutions of the single-particle (microscale)
problem~(\S\ref{sec:solution}). Asymptotic coefficients from solutions
of the microscale problem capture far-field decays of perturbations to
the equilibrium state. These are used to calculate bulk ion fluxes
in~\S\ref{sec:fluxes}, and to derive an average momentum equation
in~\S\ref{sec:momentum}. Results are presented
in~\S\ref{sec:efieldalone} for composites with negatively charged
inclusions in polymer gels saturated with NaCl electrolyte. The
subsections therein examine the incremental pore
mobility~(\S\ref{sec:incporevel1}), electroosmotic
pumping~(\S\ref{sec:incporevel2}), incremental pressure
gradient~(\S\ref{sec:incporevel3}) and, finally, species
fluxes~(\S\ref{sec:specfluxes}) and electrical
conductivity~(\S\ref{sec:eleccond}). A brief summary follows in
\S\ref{sec:summary}.

\section{Theory} \label{sec:theory}

\begin{figure}
  \begin{center}
    \input{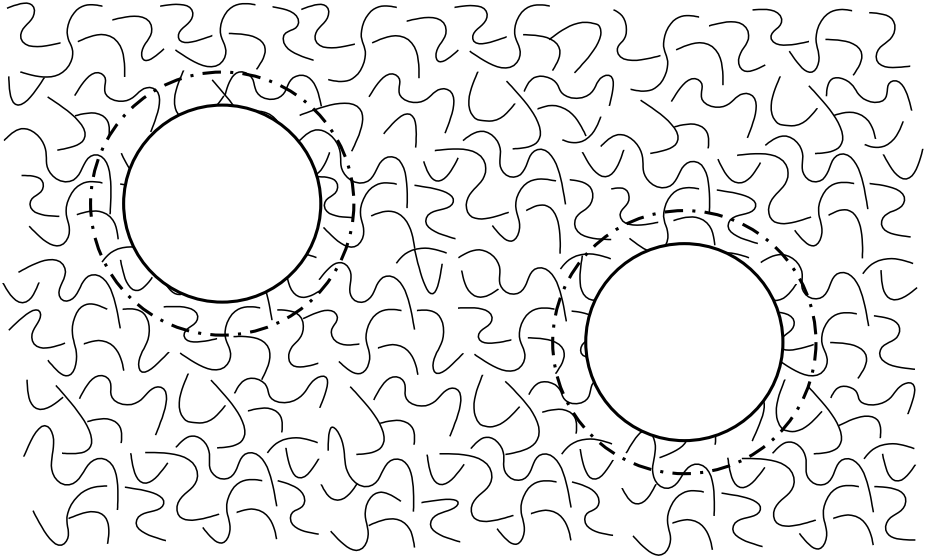}
    \caption{\label{fig:figure1} Schematic of the {\em microscale}
      system under consideration. Charged, impenetrable inclusions
      ({\em solid circles}) with radius $a \sim 10$~nm--$10~\mu$m are
      embedded in a continuous polymer gel ({\em solid filaments})
      saturated with aqueous electrolyte. Diffuse double layers ({\em
      dash-dotted circles}) with thickness $\kappa^{-1} \sim
      1$--100~nm are perturbed by the application of an average
      electric field $-\langle \grad \psi \rangle$, pressure gradient
      $\langle \grad p \rangle$, or electrolyte concentration gradient
      $\langle \grad n_j \rangle$. The Brinkman screening length $\bsl
      \sim 1$--10~nm that specifies the Darcy permeability $\bsl^2$ of
      the gel is often small compared to the radius of the
      inclusions.}
  \end{center}
\end{figure}

The microstructure of the composites considered in this work is
depicted in figure~\ref{fig:figure1}. The continuous phase is a porous
medium comprised of an electrically neutral, electrolyte-saturated
polymer gel\footnote{A {\em gel} refers to a network of polymer chains
that is cross-linked so as to exhibit a solid-like (elastic) response
to an applied stress.}. Polyacrylamide gels are routinely used for the
electrophoretic separation of DNA segments in aqueous media. Their
porosity may be controlled by adjusting the average densities and
ratio of the monomer (acrylamide) and cross-linker.  In this work, the
hydrodynamic permeability is characterized by the Darcy permeability
$\bsl^2$ (square of the Brinkman screening length), which reflects the
hydrodynamic size $a_s$ and concentration $n_s$ of the polymer
segments. In turn, these reflect the degree of cross-linking and the
affinity of the polymer for the solvent.

Embedded in the polymer are randomly dispersed spherical
inclusions. In model systems, the inclusions are envisioned to be
monodisperse silica or polymeric spheres, which typically have radii
in the range $a = 10$~nm--$10~\mu$m and bear a surface charge when
dispersed in aqueous media. The surface charge density may vary with
the bulk ionic strength and pH of the electrolyte. In this work,
however, the surface charge is to be inferred from the bulk ionic
strength and surface potential $\zeta$. Because the inclusions are
impenetrable with zero surface capacitance and conductivity, no-flux
and no-slip boundary conditions apply at their surfaces.

Note that the mobile ions whose charge is opposite to the
surface-bound immobile charge are referred to as {\em counter-ions},
with the other species referred to as {\em co-ions}. For simplicity,
the counter-charges, \ie, the dissociated counter-ions, are assumed
indistinguishable from the electrolyte counter-ions. Surrounding each
inclusion is a diffuse layer of mobile charge, with Debye thickness
$\kappa^{-1}$ and excess of counter-ions. As described below, the
layer structure is calculated from the well-known Poisson-Boltzmann
equation.

In this work, the polymer gel is assumed not to hinder ion motion. For
larger ions and dense polymer gels, the influence of the network on
the diffusive, electromigrative and convective fluxes may be modeled
with an equation of motion for an ion
\begin{equation} \label{eqn:ionmotion}
  \gamma (\vect{u} - \vect{v}) - \gamma' \vect{v} + \vect{f} =
  \vect{0},
\end{equation}
where the first term represents the hydrodynamic drag due to relative
motion; $\vect{u}$ and $\vect{v}$ are the (average) fluid and ion
velocities, and $\gamma$ is the friction coefficient. The second term
approximates the force exerted by the polymer gel on the ion, with the
friction coefficient $\gamma'$ reflecting the relative size of the ion
and polymer interstices.  The third term accounts for electrical and
thermal (Brownian) forces, depending on the time-scale of
interest. The unhindered ion velocity is $\vect{v}^0 = \vect{u} +
\vect{f} / \gamma$, so the hindered velocity may be written $\vect{v}
= \vect{v}^0 \gamma / (\gamma + \gamma')$. Therefore, under steady
conditions, the ion conservation equation $\dive{(n \vect{v})} = 0$ is
independent of $\gamma'$ ($n$ is the ion number density). When
$\vect{u} = \vect{0}$, for example, the ion will diffuse (or
electromigrate) with an {\em effective} diffusivity (or mobility) $D^e
= D \gamma / (\gamma + \gamma')$, where $D$ is the unhindered
diffusivity. It follows that $\vect{v} = \vect{v}^0 D^e / D$, so the
hindered flux equals the unhindered flux multiplied by the ratio of
the hindered to unhindered ion diffusivities (mobilities).

Now consider the influence of hindered ion migration on the fluid
momentum conservation equation. From Eqn.~(\ref{eqn:ionmotion}), the
(hydrodynamic drag) force exerted by an ion on the solvent is
\begin{equation}
  \gamma (\vect{v} - \vect{u}) = (\vect{f} - \gamma' \vect{u}) \gamma
  / (\gamma + \gamma'),
\end{equation}
where $\vect{f} = - z e \grad \psi$ is the electrical force on the ion
($z$ is the valance, $e$ is the elementary charge, and $\psi$ is the
electrostatic potential). It follows that the net force (per unit
volume) exerted by the ions on the fluid is
\begin{equation}
  - \sum_{j=1}^{N} n_j (\gamma'_j \vect{u} + z_j e \grad{\psi}) D^e_j
  / D_j,
\end{equation}
where the sum is over all $N$ ion species. Clearly, as $\gamma'_j /
\gamma_j \rightarrow 0$ when the hindrance of the polymer is
negligible, ions transfer their electrical force to the fluid. As
$\gamma'_j / \gamma_j \rightarrow \infty$, however, immobilized ions
transfer the electrical force to the polymer, so the net force exerted
by each ion on the fluid becomes $- \gamma_j \vect{u}$.

\begin{figure}
  \begin{center}
    \input{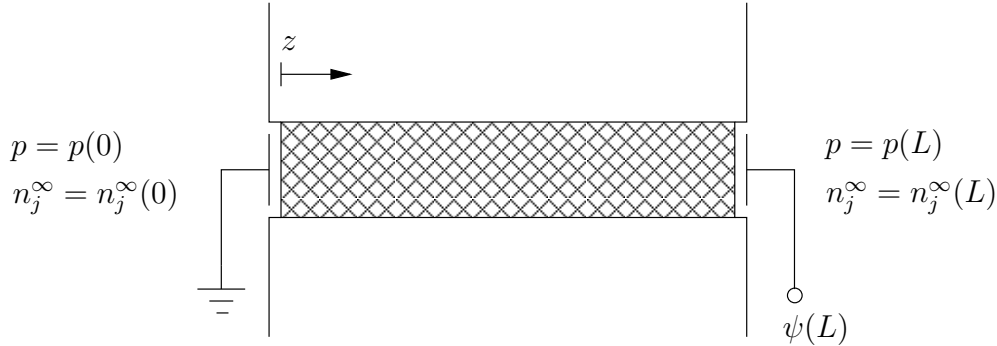}
    \caption{\label{fig:figure2} Schematic of the {\em macroscale}
      system under consideration. A polymer gel embedded with
      spherical charged inclusions (see figure~\ref{fig:figure1})
      separates (by length $L$) two reservoirs containing electrolyte
      with different species concentrations ($n_j^{\infty}(z=0)$ and
      $n_j^{\infty}(z=L)$) and, possibly, pressures ($p(z=0)$ and
      $p(z=L)$). Electrodes on each side of the `bridge' permit an
      electric field to be applied and the differential electrostatic
      potential $\Delta \psi = \psi(z=L) - \psi(z=0)$ to be
      measured. The walls of the bridge are impenetrable and
      non-conducting.}
  \end{center}
\end{figure}

A simple apparatus to realize the conditions under which the theory
may be applied is depicted in figure~\ref{fig:figure2}. The composite
bridges two reservoirs, each, in general, with a different electrolyte
concentration and pressure. Electrodes are placed at each end of the
bridge, so either a uniform electric field can be established or an
average electric field strength measured. The channel is to realize
constant ion fluxes under steady or quasi-steady conditions. This
makes the averaged microscale transport equations easier to solve,
but, in general, the averaged equations apply to macroscale fluxes in
two- and three-dimensional geometries when the (average) inclusion
number density, ion concentrations, electric field and fluid velocity
vary slowly in space and time.

\subsection{The electrokinetic transport equations} \label{sec:equations}

The transport equations and boundary conditions are presented here in
dimensional form. They comprise the non-linear Poisson-Boltzmann
equation
\begin{equation} \label{eqn:pbeqn}
  \epsilon_o \epsilon_s \lapl \psi = - \sum_{j=1}^{N} (n_j - n^f_j)
  z_j e,
\end{equation}
where $\epsilon_o$ and $\epsilon_s$ are the permittivity of a vacuum
and dielectric constant of the electrolyte; $n_j$ are the
concentrations of the $j$th mobile ions with valences $z_j$; and
$\psi$ and $e$ are the electrostatic potential and elementary charge.
In this work, the polymer is uncharged, so the fixed charge density
$n_j^f$ is zero\footnote{For convenience, the valence of the fixed
charge is set opposite to that of its respective (mobile) counter-ion
in Eqn.~(\ref{eqn:pbeqn}); this defines the concentration $n^f_j$.}.

Transport of the mobile ions is governed by
\begin{equation} \label{eqn:iontransport}
  6 \pi \eta a_j (\vect{u} - \vect{v}_j) - z_j e \grad \psi - \kb T
  \grad \ln{n_j} = 0 \ \ (j=1,...,N),
\end{equation}
where $a_j$ are Stokes radii of the ions, obtained from limiting
conductances or diffusivities; $\eta$ is the electrolyte viscosity;
$\vect{u}$ and $\vect{v}_j$ are the fluid and ion velocities; and $\kb
T$ is the thermal energy.

Ion diffusion coefficients, which are adopted throughout this paper,
are
\begin{equation} \label{eqn:diffusivity}
  D_j = \kb T / (6 \pi \eta a_j).
\end{equation}

As usual, the double-layer thickness (Debye length)
\begin{equation} \label{eqn:kappa}
  \kappa^{-1} = \sqrt{\kb T \epsilon_s \epsilon_o / (2 I e^2)}
\end{equation}
emerges from Eqns.~(\ref{eqn:pbeqn}) and~(\ref{eqn:iontransport})
where,
\begin{equation} \label{eqn:ionicstrength}
  I = (1/2) \sum_{j=1}^{N} z^2_j n^\infty_j
\end{equation}
is the bulk (average) ionic strength, with $n_j^\infty$ the bulk ion
concentrations. 

Ion conservation demands
\begin{equation} \label{eqn:ionconservation}
  \partial n_j / \partial t = -\dive (n_j \vect{v}_j) \ \ (j=1,...,N),
\end{equation}
where $t$ is the time, and the ion fluxes
\begin{equation}
  \vect{j}_j = n_j \vect{v}_j = -D_j \grad n_j - z_j e \frac{D_j}{\kb
  T} n_j \grad \psi + n_j \vect{u}
\end{equation}
are obtained from Eqn.~(\ref{eqn:iontransport}).

Similarly, momentum and mass conservation require
\begin{equation} \label{eqn:linnseqns}
  \rho_s \partial \vect{u} / \partial t = \eta \lapl \vect{u} - \grad
  p - (\eta / \bsl^2) \vect{u} - \sum_{j=1}^{N} n_j z_j e \grad \psi
\end{equation}
and
\begin{equation} \label{eqn:incomp}
  \dive \vect{u} = 0,
\end{equation}
where $\rho_s$ and $\vect{u}$ are the electrolyte density and
velocity, and $p$ is the pressure. Note that $-(\eta / \bsl^2)
\vect{u}$ represents the hydrodynamic drag force exerted by the
polymer on the electrolyte. The Darcy permeability $\bsl^2$ (square of
the Brinkman screening length) of the gel may be expressed as
\begin{equation} \label{eqn:brinkmanscreeninglength}
  \bsl^2 = 1 / [n_s(r) 6 \pi a_s F_s] = 2 a_s^2 / [9 \phi_s(r) F_s(\phi_s)],
\end{equation}
where $n_s(r)$ is the concentration of Stokes resistance centers, with
$a_s$ and $F_s(\phi_s)$ the Stokes radius and drag coefficient of the
polymer segments. In this work, $n_s$ is constant, but, in general,
may vary with radial position $r$ from the center of each
inclusion. Note also that the Brinkman screening length is adjusted
according to Eqn.~(\ref{eqn:brinkmanscreeninglength}) by varying the
(uniform) polymer segment density with Stokes radius $a_s =
1$~\AA. The drag coefficient $F_s$ is obtained from a correlation
developed by~\cite{Koch:1999} for random fixed beds of spheres. While
the microstructure of a polymer gel is clearly not the same as that of
a random bed of spheres, the model is intended to capture the
significant influence of hydrodynamic interactions on the permeability
when the volume fraction of polymer is not small. In this work,
however, only the reported values of $\bsl$ are relevant. For example,
the value $\bsl \approx 0.951$~nm, which is adopted for the principal
set of results tabulated below, reflects a polymer segment
concentration $n_s$ that yields $\bsl = 1$~nm according to
Eqn.~(\ref{eqn:brinkmanscreeninglength}) when the Stokes radius $a_s =
1$~\AA \ {\em and} the drag coefficient $F_s = 1$. Because the
hydrodynamic volume fraction $\phi_s = n_s (4/3) \pi a_s^3 > 0$,
$F_s(\phi_s) > 1$ and, hence, $\bsl$ is slightly less than the
targeted value.

\subsubsection{Inner (particle surface) boundary conditions}

Either the equilibrium surface potential $\zeta$ or surface charge
density $\sigma$ may be specified. Because the surface ($r = a$) is
assumed impenetrable with zero capacitance and conductivity, the
surface charge is constant, permitting no-flux boundary conditions for
each (mobile) ion species. As usual, the no-slip boundary condition
applies. It follows that (inner) boundary conditions are either
\begin{equation} \label{eqn:zetapotential}
  \psi = \zeta \mbox{ at } r = a
\end{equation}
or
\begin{equation} \label{eqn:psibc1}
  \epsilon_s \epsilon_o \grad \psi |_{out} \cdot \hat{\vect{n}} -
  \epsilon_p \epsilon_o \grad \psi |_{in} \cdot \hat{\vect{n}} = -
  \sigma \mbox{ at } r = a,
\end{equation}
with
\begin{equation} \label{eqn:impenetrable}
  n_j \vect{v}_j \cdot \hat{\vect{n}} = 0 \mbox{ at } r = a
\end{equation}
and
\begin{equation} \label{eqn:noslip}
  \vect{u} = \vect{0} \mbox{ at } r = a,
\end{equation}
where $\hat{\vect{n}} = \vect{e}_r$ is an outward unit normal and
$\epsilon_p$ is the particle dielectric constant.

\subsubsection{Outer (far-field) boundary conditions}

Neglecting particle interactions requires far-field boundary
conditions
\begin{equation} \label{eqn:psibc2}
  \psi \rightarrow - \vect{E} \cdot \vect{r} \mbox{ as } r \rightarrow
  \infty,
\end{equation}
\begin{equation} \label{eqn:averageconc}
  n_j \rightarrow n_j^\infty + \vect{B}_j \cdot \vect{r} \mbox{ as } r
  \rightarrow \infty,
\end{equation}
and
\begin{equation} \label{eqn:rest}
  \vect{u} \rightarrow \vect{U} \mbox{ as } r \rightarrow \infty,
\end{equation}
where $\vect{E}$, $\vect{B}_j$ and $\vect{U}$ are, respectively, a
constant electric field, constant species concentration gradients, and
constant far-field velocity.

\subsection{Solution of the equations} \label{sec:solution}

\subsubsection{Equilibrium state}

When $\vect{E} = \vect{B}_j = \vect{U} = \vect{0}$, equilibrium is
specified according to
\begin{equation}
  \epsilon_o \epsilon_s \lapl \psi^0 = - \sum_{j=1}^{N} (n^0_j -
  n^f_j) z_j e,
\end{equation}
\begin{equation}
  0 = \dive [D_j \grad{n^0_j} + z_j e \frac{D_j}{\kb T} n^0_j
    \grad{\psi^0}]
\end{equation}
and
\begin{equation}
  0 = - \grad p^0 - \sum_{j=1}^{N} n^0_j z_j e \grad \psi^0,
\end{equation}
with boundary conditions
\begin{equation}
  \psi^0 = \zeta \mbox{ at } r = a,
\end{equation}
\begin{equation}
  \epsilon_s \epsilon_o \grad \psi^0 |_{out} \cdot \vect{e}_r -
  \epsilon_p \epsilon_o \grad \psi^0 |_{in} \cdot \vect{e}_r = -
  \sigma \mbox{ at } r = a,
\end{equation}
\begin{equation}
  n^0_j [D_j \grad{n^0_j} + z_j e \frac{D_j}{\kb T} n^0_j
  \grad{\psi^0}] \cdot \vect{e}_r = 0 \mbox{ at } r = a,
\end{equation}
and
\begin{equation}
  \psi^0 \rightarrow 0 \mbox{ as } r \rightarrow \infty,
\end{equation}
\begin{equation}
  n^0_j \rightarrow n_j^\infty \mbox{ as } r \rightarrow \infty.
\end{equation}

\subsubsection{Linearized perturbed state}

Perturbations to the equilibrium state (above) are introduced via
\begin{equation}
  \psi = \psi^0 - \vect{E} \cdot \vect{r} + \psi',
\end{equation}
\begin{equation}
  n_j = n_j^0 + \vect{B}_j \cdot \vect{r} + n'_j,
\end{equation}
and
\begin{equation}
  p = p^0 + \vect{P} \cdot \vect{r} + p',
\end{equation}
where the first terms on the right-hand sides denote the equilibrium
values, with the primed quantities denoting perturbations. Note that
$\vect{P}$ is the far-field pressure gradient required to sustain a
far-field velocity $\vect{U} = - (\bsl^2 / \eta) \vect{P}$.

With `forcing'
\begin{equation}
  \vect{X} = X \vect{e}_z,
\end{equation}
where $X \in \{E, B_j, U\}$\footnote{In general, a linear combination
  of these variables.}, the linearized perturbations are symmetric
  about the $z$-axis ($\theta = 0$) of a spherical polar coordinate
  system, taking the forms
\begin{equation} \label{eqn:pert1}
  \psi' = \hat{\psi}(r) \vect{X} \cdot \vect{e}_r
\end{equation}
\begin{equation}  \label{eqn:pert2}
  n_j' = \hat{n}_j(r) \vect{X} \cdot \vect{e}_r
\end{equation}
and
\begin{equation}
  \vect{u} = \vect{U} + \vect{u}',
\end{equation}
where\footnote{Equation~(\ref{eqn:velocity}) guarantees a solenoidal
(incompressible) velocity field, which permits the momentum equation
to be solved by applying the curl $\grad \times$, thereby eliminating
the pressure and yielding a scalar equation for the single non-zero
component of the vorticity $\grad \times \vect{u} = \omega_\phi
\vect{e}_\phi$.}
\begin{eqnarray}\label{eqn:velocity}
  \vect{u}' &=& \grad \times \grad \times h(r) \vect{X} \nonumber \\
  &=& - 2 (h_r / r) (\vect{X} \cdot \vect{e}_r) \vect{e}_r - (h_{rr} +
  h_{r} / r) (\vect{X} \cdot \vect{e}_\theta) \vect{e}_\theta.
\end{eqnarray}

The perturbations satisfy
\begin{equation}
  \epsilon_o \epsilon_s \lapl \psi' = - \sum_{j=1}^{N} (n'_j +
  \vect{B}_j \cdot \vect{r}) z_j e,
\end{equation}
and
\begin{equation}
  \dive \vect{j}_j = 0,
\end{equation}
where
\begin{eqnarray}
  \vect{j}_j = - D_j (\grad{n'_j} + \vect{B}_j) - z_j e \frac{D_j}{\kb
  T} (n_j' + \vect{B}_j \cdot \vect{r}) \grad{\psi^0} \nonumber \\ -
  z_j e \frac{D_j}{\kb T} n^0_j (\grad{\psi'}-\vect{E}) + n^0_j
  (\vect{U} + \vect{u}'),
\end{eqnarray}
and
\begin{eqnarray}
  \eta \lapl \vect{u}' - \grad p' - (\eta / \bsl^2) (\vect{U} +
  \vect{u}') - \sum_{j=1}^{N} n^0_j z_j e (\grad \psi' - \vect{E})
  \nonumber \\ - \sum_{j=1}^{N} (n'_j + \vect{B}_j \cdot \vect{r}) z_j
  e \grad \psi^0 = 0
\end{eqnarray}
\begin{equation}
  \dive \vect{u}' = 0,
\end{equation}
with boundary conditions
\begin{equation} \label{eqn:perturbedfieldbc}
  \epsilon_s \epsilon_o (\grad \psi' - \vect{E}) |_{out} \cdot
  \vect{e}_r - \epsilon_p \epsilon_o (\grad \psi' - \vect{E}) |_{in}
  \cdot \vect{e}_r = 0 \mbox{ at } r = a,
\end{equation}
\begin{eqnarray}
  [D_j (\grad{n'_j} + \vect{B}_j) + z_j e
    \frac{D_j}{\kb T} (n'_j + \vect{B}_j \cdot \vect{r}) \grad{\psi^0}
    \nonumber \\ + z_j e \frac{D_j}{\kb T} n^0_j
    (\grad{\psi'}-\vect{E}) - n^0_j (\vect{U} + \vect{u}')] \cdot
  \vect{e}_r = 0 \mbox{ at } r = a,
\end{eqnarray}
\begin{equation}
  \vect{u}' = -\vect{U} \mbox{ at } r = a,
\end{equation}
and
\begin{equation} \label{eqn:farfield1}
  \psi' \rightarrow (\vect{X} \cdot \vect{e}_r) D^X / r^2 \mbox{ as }
  r \rightarrow \infty,
\end{equation}
\begin{equation} \label{eqn:farfield2}
  n'_j \rightarrow (\vect{X} \cdot \vect{e}_r) C^X_j / r^2 \mbox{ as }
  r \rightarrow \infty,
\end{equation}
\begin{equation}  \label{eqn:farfield3}
  \vect{u}' \rightarrow - 2 (C^X / r^3) (\vect{X} \cdot \vect{e}_r)
  \vect{e}_r - (C^X/r^3) (\vect{X} \cdot \vect{e}_\theta)
  \vect{e}_\theta \mbox{ as } r \rightarrow \infty.
\end{equation}

In the far field, the velocity disturbance $\vect{u}'$ is proportional
to the gradient of $p'$, which, like the electrostatic potential and
ion concentrations, is dipolar.  Accordingly, $\vect{u}'$ decays as
$r^{-3}$, and $D^X$ and $C_j^X$ will often be referred to as the
strength of the electrostatic and concentration polarization (or
dipole moments) induced by the field $X \in \{E, B_j, U\}$.

The dimensions of the {\em asymptotic coefficients} $D^X$, $C_j^X$ and
$C^X$, which depend on the respective $X \in \{E, B_j, U\}$, are
easily worked out by inspecting
Eqns.~(\ref{eqn:farfield1})--(\ref{eqn:farfield3}). For convenience,
dimensionless values are presented in the tables below with $a$, $u^*
= \epsilon_s \epsilon_o (\kb T / e)^2 / (\eta a)$, $2 I$ and $\kb T /
e$ as the scales for length, velocity, ion concentrations, and
electrostatic potential, respectively.

\subsubsection{Superposition}

The equations are solved using a numerical methodology developed
by~\cite{Hill:2003a} for the electrophoretic mobility of
polymer-coated colloids. Solutions with $E$, $B_j$ and $U$ set to
arbitrary values can be computed, provided $\sum_{j=1}^N z_j B_j = 0$
to ensure an electrically neutral far-field. However, when $N$ species
are assembled into $M$ electroneutral groups (\eg, electrolytes or
neutral tracers), each with far-field gradient $B_k$ ($k = 1,...,M$),
it is expedient to compute solutions with only one non-zero value of
$E$, $B_k$ or $U$. Then, arbitrary solutions can be obtained by linear
superposition~\citep{OBrien:1978}.

An index $k'$ is required to identify the (electroneutral) group to
which the $j$th species under consideration is assigned. Careful
consideration of the electrolyte composition and ion valences is
required to ensure consistency. For $z$-$z$ electrolytes it is
convenient to set $B_j = B_k$, whereas for a single 2-1 electrolyte
(\eg, CaCl$_2$) is it satisfactory to set $B_j = B_{k'} / |z_j|$. For
the relatively simple situations considered in this work, the (single)
electroneutral group is NaCl, so $M = 1$ with $k = k' = 1$, and $j =
1$ and 2 for Na$^+$ and Cl$^-$, respectively.

Note that $C_j^{B_k}$, for example, is the asymptotic coefficient for
the perturbed {\em concentration} of the $j$th species induced by the
$k$th concentration gradient $B_k$, whereas $C^{B_k}$ (without a
subscript) denotes the asymptotic coefficient for the {\em flow}
induced by $B_k$.

For neutral species, the concentration disturbance produced by a
single impenetrable sphere yields $C_j^{B_{k'}} = (1/2) a^3$,
otherwise $C_j^{B_{k}} = 0$ ($k \ne k'$). Clearly, the asymptotic
coefficients for charged species, whose concentration perturbations
are influenced by electromigration, are not the same as for neutral
species; for ions, $C_j^{B_{k'}} \rightarrow (1/2) a^3$ as $|\zeta|
\rightarrow 0$, however.

With co-linear forcing and bulk electroneutrality, linear
superposition gives far-field decays
\begin{equation}
  \psi' \rightarrow (1 / r^2)[E D^E + \sum_{k=1}^{M} B_k D^{B_k} + U
    D^U] (\vect{e}_z \cdot \vect{e}_r) \mbox{ as } r \rightarrow
  \infty,
\end{equation}
\begin{equation}
  n_j' \rightarrow (1 / r^2)[E C_j^E + \sum_{k=1}^{M} B_k C_j^{B_k} +
    U C_j^U] (\vect{e}_z \cdot \vect{e}_r) \mbox{ as } r \rightarrow
  \infty,
\end{equation}
and
\begin{eqnarray} 
  \vect{u}' \rightarrow - (2 / r^3) [E C^E + \sum_{k=1}^{M} B_k
    C^{B_k} + U C^U] (\vect{e}_z \cdot \vect{e}_r) \vect{e}_r
    \nonumber \\ - (1/r^3) [E C^E + \sum_{k=1}^{M} B_k C^{B_k} + U
    C^U] (\vect{e}_z \cdot \vect{e}_\theta) \vect{e}_\theta \mbox{ as
    } r \rightarrow \infty.
\end{eqnarray}

This work is {\em primarily} concerned with situations where only one
$X \in \{E, B_j, U\}$ is applied. Following \cite{OBrien:1978}, these
are referred to as the (E), (B) and (U) (microscale)
problems. Algebraic or differential relationships between the averaged
fields may be applied to ensure zero average current density, for
example. The next section relates (microscale) $\vect{E}$,
$\vect{B}_j$ and $\vect{U}$ to the averaged (macroscale) fields, \eg,
$-\langle \grad{\psi}\rangle$, $\langle \grad{n_j} \rangle$ and
$\langle \vect{u} \rangle$, in {\em dilute} composites.

\subsection{Averaged (bulk) fluxes} \label{sec:fluxes}

Here we calculate the average flux of the $j$th species
\begin{equation} \label{eqn:average}
  \langle \vect{j}_j \rangle = V^{-1} \int \vect{j}_j \mbox{d}V,
\end{equation}
where the volume of integration includes the continuous and discrete
phases. If the size of the {\em representative elementary volume} is
between the micro- and macro-scales, the result is equivalent to
sampling the flux (at a point) over all micro-structural
configurations (ensemble average).

Following \cite{Saville:1979} and \cite{OBrien:1981}, the averaging
can be accomplished by adding and subtracting the flux
\begin{equation}
  -D_j \grad n_j - z_j e \frac{D_j}{\kb T} n_j^\infty \grad \psi +
  n_j^\infty \vect{u}
\end{equation}
from the integrand in Eqn.~(\ref{eqn:average}). This yields the
macroscopic electromigrative, diffusive and convective fluxes in the
absence of inclusions, plus an integral whose integrand is
exponentially small beyond the diffuse double layers, \ie,
\begin{eqnarray} \label{eqn:integral}
  \langle \vect{j}_j \rangle = - z_j e \frac{D_j}{\kb T} n_j^\infty
  \langle \grad{\psi} \rangle - D_j \langle \grad{n_j} \rangle +
  n_j^\infty \langle \vect{u} \rangle \nonumber \\ + V^{-1} \int [ z_j
  e \frac{D_j}{\kb T} n_j^\infty \grad{\psi} + D_j \grad{n_j} -
  n_j^\infty \vect{u} + \vect{j}_j ] \mbox{d}V.
\end{eqnarray}
Applying the divergence theorem\footnote{Note also that $\int_V
  \vect{\alpha} \mbox{d}V = \int_A \vect{x} \vect{\alpha} \cdot
  \hat{\vect{n}} \mbox{d}A - \int_V \vect{x} \dive{\vect{\alpha}}
  \mbox{d}V$, where $\vect{\alpha}$ represents an arbitrary vector
  field.} and noting that $\dive \vect{j}_j = 0$, the volume integral
  in Eqn.~(\ref{eqn:integral}) becomes
  \begin{eqnarray}
    \int z_j e \frac{D_j}{\kb T} n_j^\infty \psi \hat{\vect{n}}
    \mbox{d}A - \int z_j e \frac{D_j}{\kb T} n_j [\grad{\psi} \cdot
    \hat{\vect{n}}] \vect{r} \mbox{d}A \nonumber \\ + \int D_j n_j
    \hat{\vect{n}} \mbox{d}A - \int D_j [\grad{n_j} \cdot
    \hat{\vect{n}}] \vect{r} \mbox{d}A \nonumber \\ + \int (n_j -
    n_j^\infty) (\vect{u} \cdot \hat{\vect{n}}) \vect{r} \mbox{d}A,
\end{eqnarray}
where the surface integrals enclose the inclusions and their
respective equilibrium double layer, with $\hat{\vect{n}}$ directed
outward, into the fluid.

For dilute composites, \ie, when $n (4/3) \pi (a + \kappa^{-1})^3 \ll
1$, the integral over a representative volume $V$ equals $nV$
integrals with a single particle at $\vect{r} = \vect{0}$, each with
$\hat{\vect{n}} = \vect{e}_r$.  Therefore, noting that $n_j^0 -
n_j^\infty$ is exponentially small as $r \rightarrow \infty$, and that
$n'_j$ and $\vect{B}_j \cdot \vect{r}$ are odd functions of position,
the fluxes become
\begin{eqnarray} \label{eqn:flux}
  \langle \vect{j}_j\rangle \approx n_j^\infty \langle \vect{u}
  \rangle - z_j e \frac{D_j}{\kb T} n_j^\infty \langle \grad{\psi}
  \rangle - D_j \langle \grad n_j \rangle \nonumber \\ + n z_j e
  \frac{D_j}{\kb T} n_j^\infty \int_{r \rightarrow \infty} [\psi'-
  (\grad{\psi'} \cdot \vect{r})] \vect{e}_r \mbox{d}A \nonumber \\ + n
  D_j \int_{r \rightarrow \infty} [n'_j - (\grad{n'_j} \cdot
  \vect{r})] \vect{e}_r \mbox{d}A.
\end{eqnarray}

Superposing solutions of the independent single-particle problems, the
microscale electromigrative and diffusive contributions to the average
flux, \ie, the last two terms in Eqn.~(\ref{eqn:flux}),
are\footnote{Note that $\vect{r} (\vect{\alpha} \cdot \vect{e}_r)$ has
been written as $(\vect{\alpha} \cdot \vect{r}) \vect{e}_r$ because
$\vect{r} = r \vect{e}_r$, where $\vect{\alpha}$ represents an
arbitrary vector field.},
\begin{eqnarray}
  n z_j e \frac{D_j}{\kb T} n_j^\infty \int_{r\rightarrow \infty}
  [\psi' - (\grad{\psi'} \cdot \vect{r})] \vect{e}_r \mbox{d}A =
  \nonumber \\ n 4 \pi z_j e \frac{D_j}{\kb T} n_j^\infty [\vect{E}
  D^E + \sum_{k=1}^{M} \vect{B}_{k} D^{B_{k}} + \vect{U} D^U],
\end{eqnarray}
and
\begin{eqnarray}
  n D_j \int_{r\rightarrow \infty} [n'_j - (\grad{n'_j} \cdot
    \vect{r})] \vect{e}_r \mbox{d}A = n 4 \pi D_j [\vect{E} C_j^E +
    \sum_{k=1}^M \vect{B}_{k} C_{j}^{B_{k}} + \vect{U} C_j^U].
\end{eqnarray}
Substituting these into Eqn.~(\ref{eqn:flux}) gives
\begin{eqnarray} \label{eqn:finalflux}
  \langle \vect{j}_j\rangle \approx n_j^\infty \langle \vect{u}
  \rangle - z_j e \frac{D_j}{\kb T} n_j^\infty \langle \grad{\psi}
  \rangle - D_j \langle \grad n_j \rangle \nonumber \\ + n 4 \pi z_j e
  \frac{D_j}{\kb T} n_j^\infty [\vect{E} D^E + \sum_{k=1}^{M}
  \vect{B}_{k} D^{B_{k}} + \vect{U} D^U] \nonumber \\ + n 4 \pi D_j
  [\vect{E} C_j^E + \sum_{k=1}^M \vect{B}_{k} C_{j}^{B_{k}} + \vect{U}
  C_j^U].
\end{eqnarray}
Note that the average fluxes are now expressed in terms of the
asymptotic coefficients from at most $2 + M$ independent
single-particle problems, each of which is solved `exactly' in this
work.

\subsection{Averaged (bulk) momentum conservation equations} \label{sec:momentum}
 
In general, an average velocity $\langle \vect{u} \rangle$ is produced
by the application of an average pressure gradient $\langle \grad p
\rangle$, electric field $-\langle \grad \psi \rangle$, or
concentration gradients $\langle \grad n_j \rangle$. This section
relates these to the asymptotic coefficients emerging from the
single-particle problem.

Averaging the fluid momentum equation gives (see Appendix~\ref{app:1})
\begin{equation}\label{eqn:avemomA}
  \vect{0} \approx - \langle \grad{p} \rangle - (\eta / \bsl^2)
  \langle \vect{u} \rangle + \eta \lapl{\vect{\langle \vect{u}
  \rangle}} - \langle \rho \grad{\psi} \rangle - n \langle \vect{f}^d
  \rangle
\end{equation}
where $\rho = \rho^0 + \rho'$ is the charge density and $\langle
\vect{f}^d \rangle$ is the average (hydrodynamic) force exerted by the
fluid on the inclusions. Note that inertia is neglected, as are
hydrodynamic and electrostatic interactions; the analysis is therefore
limited to small volume fractions $n (4/3) \pi (a + \kappa^{-1})^3 \ll
1$.

Similarly to the average fluxes, let us adopt the single-particle
problem to evaluate $\langle \vect{f}^d \rangle$. For a {\em single}
inclusion in an {\em unbounded} polymer gel,
\begin{eqnarray} \label{eqn:force} 
  \langle \vect{f}^d \rangle &\approx& \int_{r=a} [-(\vect{P} \cdot
    \vect{r} + p') \vect{\delta} + 2 \eta \vect{e}] \cdot \vect{e}_r
    \mbox{d}A \nonumber \\ &=& \int_{r \rightarrow \infty} [-
    (\vect{P} \cdot \vect{r} + p') \vect{\delta} + 2 \eta \vect{e}]
    \cdot \vect{e}_r \mbox{d}A - \int_{r=a}^{\infty} [(\eta / \bsl^2)
    \vect{u} + \rho \grad{\psi}] \mbox{d}V,
\end{eqnarray}
where $\vect{e} = (1/2) [\grad \vect{u} + (\grad \vect{u})^T]$ and
$\vect{\delta}$ is the identify tensor. Since $\dive{\vect{u}} = 0$,
$\vect{u}(r=a) = \vect{0}$, and $\grad{p} = - (\eta / \bsl^2)
\vect{u}$ and $\vect{u}' \sim r^{-3}$ as $r \rightarrow \infty$,
Eqn.~(\ref{eqn:force}) can be written
\begin{equation} \label{eqn:force1}
  \langle \vect{f}^d \rangle \approx - \int_{r \rightarrow \infty} [p'
  - (\grad{p'} \cdot \vect{r})] \vect{e}_r \mbox{d}A -
  \int_{r=a}^{\infty} \rho \grad{\psi} \mbox{d}V.
\end{equation}
Beyond the double layer,
\begin{equation}
  p' = -(\eta / \bsl^2) \int_{r}^{\infty} (2 / {r'}^3) C^X (\vect{X}
  \cdot \vect{e}_r) \mbox{d}r' = -(1 / r^2) (\eta / \bsl^2) C^X
  (\vect{X} \cdot \vect{e}_r),
\end{equation}
and because the integrand of the volume integral in
Eqn.~(\ref{eqn:force1}) is exponentially small there,
\begin{equation}
\langle \rho \grad{\psi} \rangle \approx n \int_{r=a}^{\infty} \rho
\grad{\psi} \mbox{d}V.
\end{equation}
Therefore,
\begin{eqnarray} \label{eqn:force2}
  \langle \vect{f}^d \rangle \approx (\eta / \bsl^2) 4 \pi [\vect{E}
    C^E + \sum_{k=1}^{M} \vect{B}_k C^{B_{k}} + \vect{U} C^U] - n^{-1}
    \langle \rho \grad{\psi} \rangle
\end{eqnarray}
and, hence, Eqn.~(\ref{eqn:avemomA}) becomes
\begin{eqnarray} \label{eqn:avemomB}
  \vect{0} \approx -\langle \grad{p} \rangle - (\eta / \bsl^2) \langle
  \vect{u} \rangle + \eta \lapl{\vect{\langle \vect{u} \rangle}} - n
  (\eta / \bsl^2) 4 \pi [\vect{E} C^E + \sum_{k=1}^{M} \vect{B}_k
    C^{B_{k}} + \vect{U} C^U].
\end{eqnarray}

Note that, in addition to $\langle \vect{f}^d \rangle$, an electrical
force $\langle \vect{f}^e \rangle$ and a mechanical-contact force
$\langle \vect{f}^m \rangle$ act on the inclusions. Accordingly,
static equilibrium requires
\begin{eqnarray} \label{eqn:force3}
  \langle \vect{f}^m \rangle &=& - \langle \vect{f}^e \rangle -
  \langle \vect{f}^d \rangle \nonumber \\ &\approx& - (\eta / \bsl^2)
  4 \pi [\vect{E} C^E + \sum_{k=1}^{M} \vect{B}_k C^{B_{k}} + \vect{U}
  C^U].
\end{eqnarray}
In the absence of charge, for example, $- \langle \vect{f}^m \rangle$
is equal to the drag force on a sphere embedded in a Brinkman medium
with viscosity $\eta$ and Darcy permeability $\bsl^2$. Indeed,
\cite{Brinkman:1947} solved this problem exactly, obtaining
\begin{equation} \label{eqn:brinkman}
  \langle \vect{f}^m \rangle = - 2 \pi \eta \vect{U} a (a / \bsl)^2 [1
    + 3 (\bsl / a) + 3 (\bsl / a)^2] \ \ (\zeta = 0, \ \phi
    \rightarrow 0),
\end{equation}
which shows that
\begin{equation}
  2 C^U / a^3 = 1 + 3 (\bsl / a) + 3 (\bsl / a)^2 \ \ (\zeta = 0, \
  \phi \rightarrow 0).
\end{equation}
Note that the drag force approaches the Stokes drag $6 \pi \eta a
\vect{U}$ as $\bsl / a \rightarrow \infty$. When $\bsl / a \rightarrow
0$, however, the drag approaches $2 \pi a^3 (\eta / \bsl^2) \vect{U}$
because the surface traction is dominated by the pressure (dipole)
arising from the outer Darcy flow: $\lapl{p'} = 0$ with $\vect{u} = -
(\bsl^2 / \eta) \grad{p'}$.

\subsection{Averaged (bulk) equations for unidirectional transport}

With {\em all} average fluxes in the $z$-direction, mass and momentum
conservation require constant $\langle \vect{u} \rangle$ and, hence,
\begin{eqnarray} \label{eqn:avemomC}
  \langle \grad{p} \rangle = - (\eta / \bsl^2) \langle \vect{u}
  \rangle - \phi (3/a^3) (\eta / \bsl^2) [\vect{E} C^E +
  \sum_{k=1}^{M} \vect{B}_k C^{B_{k}} + \vect{U} C^U].
\end{eqnarray}
Similarly, the (steady) average species conservation equations
$\dive{\langle \vect{j}_j \rangle} = 0$ require constant average
fluxes
\begin{eqnarray}  \label{eqn:finalfluxB}
  \langle \vect{j}_j \rangle = n_j^\infty \langle \vect{u} \rangle -
  z_j e \frac{D_j}{\kb T} n_j^\infty \langle \grad{\psi} \rangle - D_j
  \langle \grad n_j \rangle \nonumber \\ + \phi (3/a^3) z_j e
  \frac{D_j}{\kb T} n_j^\infty [\vect{E} D^E + \sum_{k=1}^{M}
  \vect{B}_{k} D^{B_{k}} + \vect{U} D^U] \nonumber \\ + \phi (3 / a^3)
  D_j [\vect{E} C_j^E + \sum_{k=1}^M \vect{B}_k C_{j}^{B_{k}} +
  \vect{U} C_j^U].
\end{eqnarray}
Note that $\dive{\langle \grad {\psi} \rangle} = 0$ in an electrically
neutral composite with uniform dielectric permittivity, so the average
electric field is also constant.

The averages can be expanded as power series in the inclusion volume
fraction \eg, $\langle \vect{u} \rangle \rightarrow \vect{U}_0 + \phi
\vect{U}_1 + O(\phi^2)$. Therefore, since the microscale equations
(asymptotic coefficients) are accurate to $O(\phi)$, the notation is
condensed by writing, for example, $\langle \vect{u} \rangle \equiv
\vect{U}$, where it is understood that $\vect{U} = \vect{U}_0 + \phi
\vect{U}_1 + O(\phi^2)$. Clearly, $\vect{E}$, $\vect{B}_j$ and
$\vect{U}$ in Eqns.~(\ref{eqn:avemomC}) and (\ref{eqn:finalfluxB})
need only to include the $O(1)$ contribution to their respective
average, \eg, $\vect{U} \rightarrow \vect{U}_0$. The following
notation is adopted for the other averaged quantities: $\vect{J}_j
\equiv \langle \vect{j}_j \rangle$, $\vect{P} \equiv \langle \grad p
\rangle$, $\vect{B}_j \equiv \langle \grad{n}_j \rangle$, $\vect{E}
\equiv - \langle \grad \psi \rangle$.

With one electrolyte ($M=1$) and, recall, bulk electroneutrality,
there are $N + 4$ independent variables ($\vect{E}, \vect{U},
\vect{P}, \vect{B}_k~(k=1), \vect{J}_j$~($j=1,...,N$)) with $N + 1$
independent equations (see Eqns.~(\ref{eqn:avemomC}) and
(\ref{eqn:finalfluxB})). Clearly, three independent variables must be
specified for a unique solution.

For clarity, the results presented below involve a 1-1 electrolyte
(NaCl), mostly with only one non-zero forcing variable. It is
important to note that, because the equations are linear, solutions
for any combination of forcing variables may be constructed. For
example, a forthcoming paper establishes the electric field strength
required to maintain a constant electrolyte flux---driven by a bulk
concentration gradient across a membrane---with zero bulk current
density.

\section{Response to an electric field} \label{sec:efieldalone}

Results are now presented for the application of an electric field in
the absence of average pressure and concentration gradients. These
conditions prevail when measuring the electrical conductivity, for
example, and they provide a relatively simple setting in which to
study the influence of inclusions on bulk electrokinetic
transport. Steady homogeneous conditions are assumed, neglecting the
influence of electrode polarization and electrochemical
reactions. Accordingly, the average velocity from
Eqn.~(\ref{eqn:avemomC}) is
\begin{eqnarray} \label{eqn:avemomE}
  \vect{U} = - \phi (3 / a^3) \vect{E} C^E + O(\phi^2),
\end{eqnarray}
and the average ion fluxes from Eqn.~(\ref{eqn:finalfluxB}) are
\begin{eqnarray} \label{eqn:fluxE}
  \vect{J}_j = z_j e \frac{D_j}{\kb T} n_j^\infty \vect{E} +
  \phi(3/a^3) z_j e \frac{D_j}{\kb T} n_j^\infty \vect{E} D^E
  \nonumber \\ + \phi (3 / a^3) D_j \vect{E} C_j^E + \phi (3/ a^3)
  n_j^\infty \vect{E} C^E + O(\phi^2).
\end{eqnarray}

Asymptotic coefficients are provided in
table~\ref{tab:efieldalonecoeffs} for a composite with Brinkman
screening length $\bsl \approx 0.951$~nm and inclusion radius $a =
100$~nm; the $\zeta$-potentials and (three) ionic strengths span
experimentally accessible ranges. With a positive electric field ($E >
0$), the counter-ions (Na$^+$) migrate toward the `front' of the
inclusions, inducing a positive electrostatic dipole moment $D^E >
0$. When the $\zeta$-potential is low, however, the dipole moment
reflects the dielectric polarization required to maintain an
impenetrable interface, so the dipole strength approaches the Maxwell
value $D^E = - (1/2) a^3$ (for non-conducting spheres) as $|\zeta|
\rightarrow 0$. The positive concentration dipole moments $C_j^E > 0$
reflect the combined influences of electromigration, diffusion, and
electroneutrality. As expected from Eqn.~(\ref{eqn:farfield3}), $C^E <
0$, because the electrical force on the fluid and, hence, the
resulting electroosmotic flow are forward ($U > 0$).

\begin{sidewaystable}
  \begin{center}
    \caption{\label{tab:efieldalonecoeffs} Dimensionless asymptotic
      coefficients (see Eqn.~(\ref{eqn:fluxE})) and {\em incremental
      pore mobility} (Eqn.~(\ref{eqn:mobility})) for bulk {\em
      electromigration} of NaCl in a Brinkman medium with charged
      spherical inclusions: $a = 100$~nm; $\bsl \approx 0.951$~nm; $T
      = 25$\degc; $D_1 \approx 1.33 \times
      10^{-9}$m$^2$s$^{-1}$~(Na$^+$); $D_2 \approx 2.03 \times
      10^{-9}$m$^2$s$^{-1}$~(Cl$^-$); $u^* = \epsilon_s \epsilon_o
      (\kb T / e)^2 / (\eta a) \approx 5.15 \times
      10^{-3}$~m~s$^{-1}$.}
    
    \begin{tabular*}{\columnwidth}{@{\extracolsep{\fill}}lllll} \hline
     
      \multicolumn{1}{c}{$\zeta e /(\kb T)$} & \multicolumn{1}{c}{$D^E
      / a^3$} & \multicolumn{1}{c}{$C_j^E \kb T/ (2 I a^3 e)$} &
      \multicolumn{1}{c}{$C^E \kb T / (u^* a^4 e)$} &
      \multicolumn{1}{c}{$U / (E \phi) = -3 C^E / a^3$}\\ & &
      \multicolumn{1}{c}{($j=1,2$)} & &
      \multicolumn{1}{c}{((nm~s$^{-1}$)/(V~cm$^{-1}$))}\\ \hline
      
$\ka=  1$ & $I=9.25\times 10^{-6}$~\M \\ \hline

$-1$ & $-3.82\times 10^{-1}$ & $1.04\times 10^{0}$ & $-1.87\times 10^{-4}$ & $1.13\times 10^{0}$ \\ 
$-2$ & $-5.71\times 10^{-2}$ & $2.04\times 10^{0}$ & $-3.66\times 10^{-4}$ & $2.20\times 10^{0}$ \\ 
$-4$ & $+8.66\times 10^{-1}$ & $3.72\times 10^{0}$ & $-6.64\times 10^{-4}$ & $3.99\times 10^{0}$ \\ 
$-6$ & $+1.60\times 10^{0}$  & $4.77\times 10^{0}$ & $-8.40\times 10^{-4}$ & $5.05\times 10^{0}$ \\ 
$-8$ & $+1.96\times 10^{0}$  & $5.28\times 10^{0}$ & $-9.04\times 10^{-4}$ & $5.44\times 10^{0}$ \\ \hline

$\ka= 10$ & $I=9.25\times 10^{-4}$~\M \\ \hline

$-1$ & $-4.74\times 10^{-1}$ & $7.30\times 10^{-2}$ & $-1.20\times 10^{-3}$ & $7.22\times 10^{0}$ \\ 
$-2$ & $-3.98\times 10^{-1}$ & $1.51\times 10^{-1}$ & $-2.42\times 10^{-3}$ & $1.46\times 10^{1}$ \\ 
$-4$ & $-1.37\times 10^{-1}$ & $3.24\times 10^{-1}$ & $-4.63\times 10^{-3}$ & $2.78\times 10^{1}$ \\ 
$-6$ & $+1.32\times 10^{-1}$ & $4.76\times 10^{-1}$ & $-5.45\times 10^{-3}$ & $3.28\times 10^{1}$ \\ 
$-8$ & $+2.95\times 10^{-1}$ & $5.64\times 10^{-1}$ & $-4.70\times 10^{-3}$ & $2.83\times 10^{1}$ \\ \hline

$\ka=100$ & $I=9.25\times 10^{-2}$~\M \\ \hline

$-1$ & $-4.96\times 10^{-1}$ & $7.83\times 10^{-3}$ & $-6.97\times 10^{-3}$ & $4.19\times 10^{1}$ \\ 
$-2$ & $-4.82\times 10^{-1}$ & $1.79\times 10^{-2}$ & $-1.42\times 10^{-2}$ & $8.51\times 10^{1}$ \\ 
$-4$ & $-4.11\times 10^{-1}$ & $5.58\times 10^{-2}$ & $-2.87\times 10^{-2}$ & $1.73\times 10^{2}$ \\ 
$-6$ & $-2.53\times 10^{-1}$ & $1.35\times 10^{-1}$ & $-3.83\times 10^{-2}$ & $2.30\times 10^{2}$ \\ 
$-8$ & $-3.81\times 10^{-2}$ & $2.43\times 10^{-1}$ & $-3.58\times 10^{-2}$ & $2.15\times 10^{2}$ \\
      
    \end{tabular*}
  \end{center}
\end{sidewaystable}

\subsection{Incremental pore velocity} \label{sec:incporevel1}

As suggested by Eqn.~(\ref{eqn:avemomE}), the ratio
\begin{equation} \label{eqn:mobility}
  U / (E \phi) = - 3 C^E / a^3,
\end{equation}
which is termed the {\em incremental pore mobility}, provides a
convenient measure of the electroosmotic pumping capacity. When
multiplied by the electric field strength and particle volume
fraction, the values in the last column of
table~\ref{tab:efieldalonecoeffs}, for example, yield the $O(\phi)$
average velocity that prevails in the absence of an applied pressure
gradient. This section examines how the strength of the flow is
related to the $\zeta$-potential and size of the inclusions, the ionic
strength of the electrolyte, and the permeability of the gel. We will
see that the pore mobility is significantly influenced by polarization
and relaxation, so the qualitative form of the relationship (with a
given gel permeability) is similar to the classical electrophoretic
mobility of dispersions~\citep{OBrien:1978}.

Consider the case in table~\ref{tab:efieldalonecoeffs} with $\ka =
100$ and $\zeta = - 1 \kb T / e$. With $E = 2$~V~cm$^{-1}$ and $\phi =
10^{-2}$, the pore velocity is $U \approx 0.84$~nm~s$^{-1}$, which is
clearly very slow. If, however, this flow is directed from a composite
with a 1~cm$^2$ cross-section into a microfluidic channel, then it is
not unreasonable to amplify the velocity by four orders of magnitude,
yielding a (modest) average velocity of 8.4~$\mu$m~s$^{-1}$.  In this
example, the permeability ($\bsl^2 = 0.951^2$~nm$^2$) is relatively
low, so higher velocities may be achieved with a comparable (weak)
electric field and inclusion volume fraction. Note that a stronger
electric field between (platinum) electrodes separated by a distance
$L \sim 5$~mm, say, produces hydrogen and oxygen bubbles. In
conventional electroosmotic pumps, much higher electric field
strengths are achieved by catalytically recombining hydrogen and
oxygen~\citep{Yao:2003b}. For the purpose of accurately determining
the pore mobility, however, higher electric field strengths are,
perhaps, unnecessary.

\begin{figure}
  \begin{center}
    \vspace{1.0cm} \includegraphics[width=5.5cm]{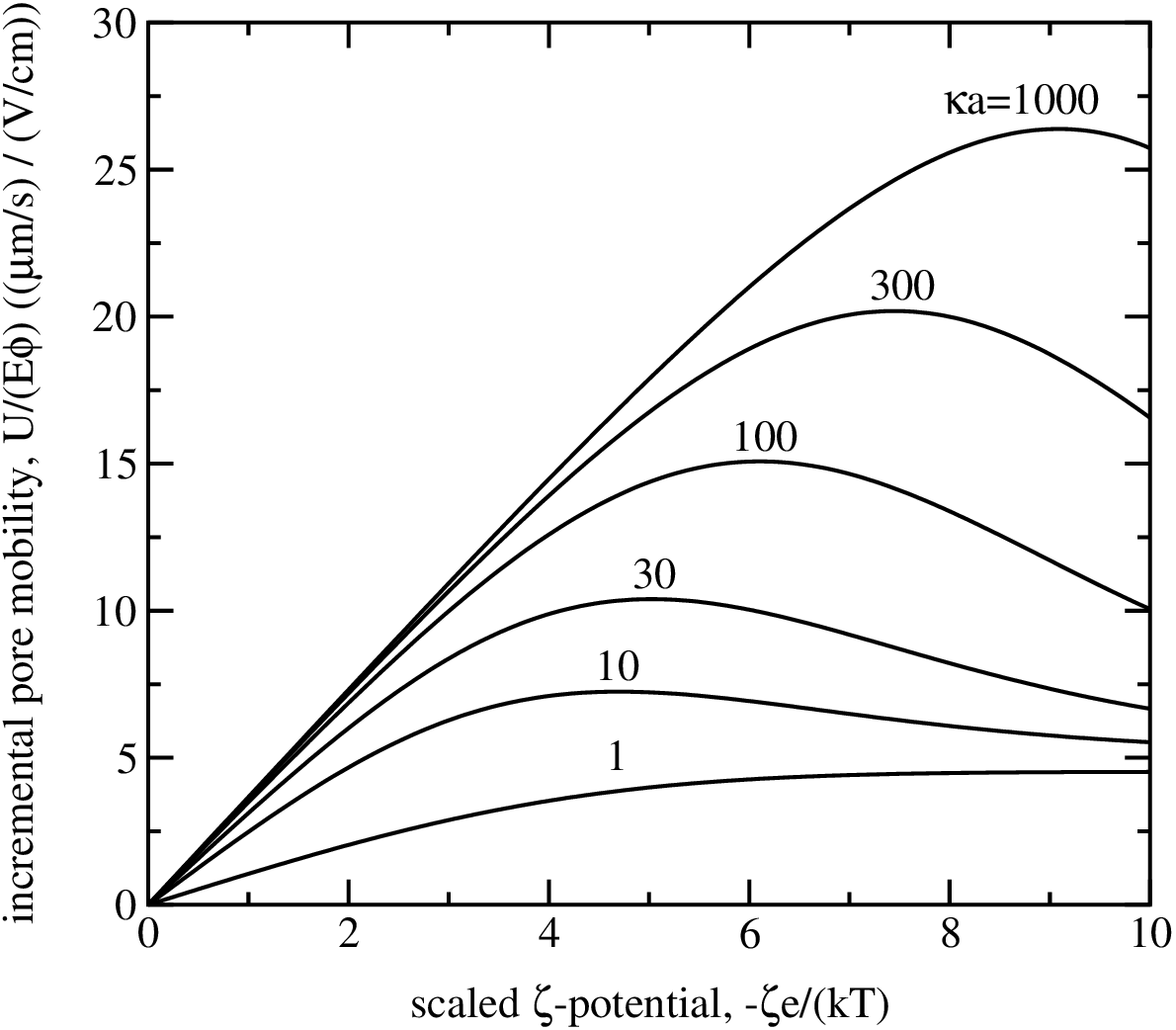}
    \hspace{0.75cm} \includegraphics[width=6cm]{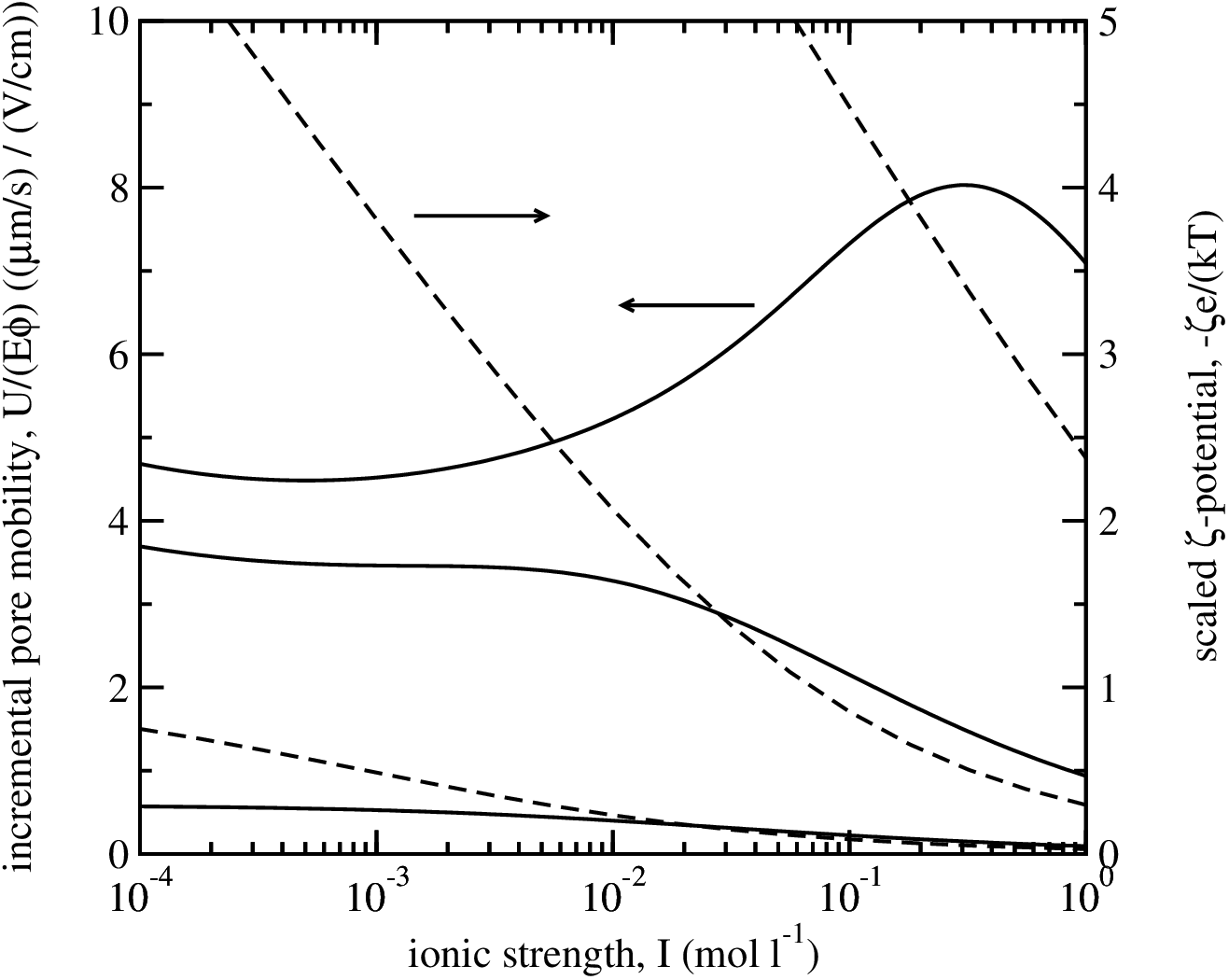}\\
    \vspace{1.0cm} \includegraphics[width=5.5cm]{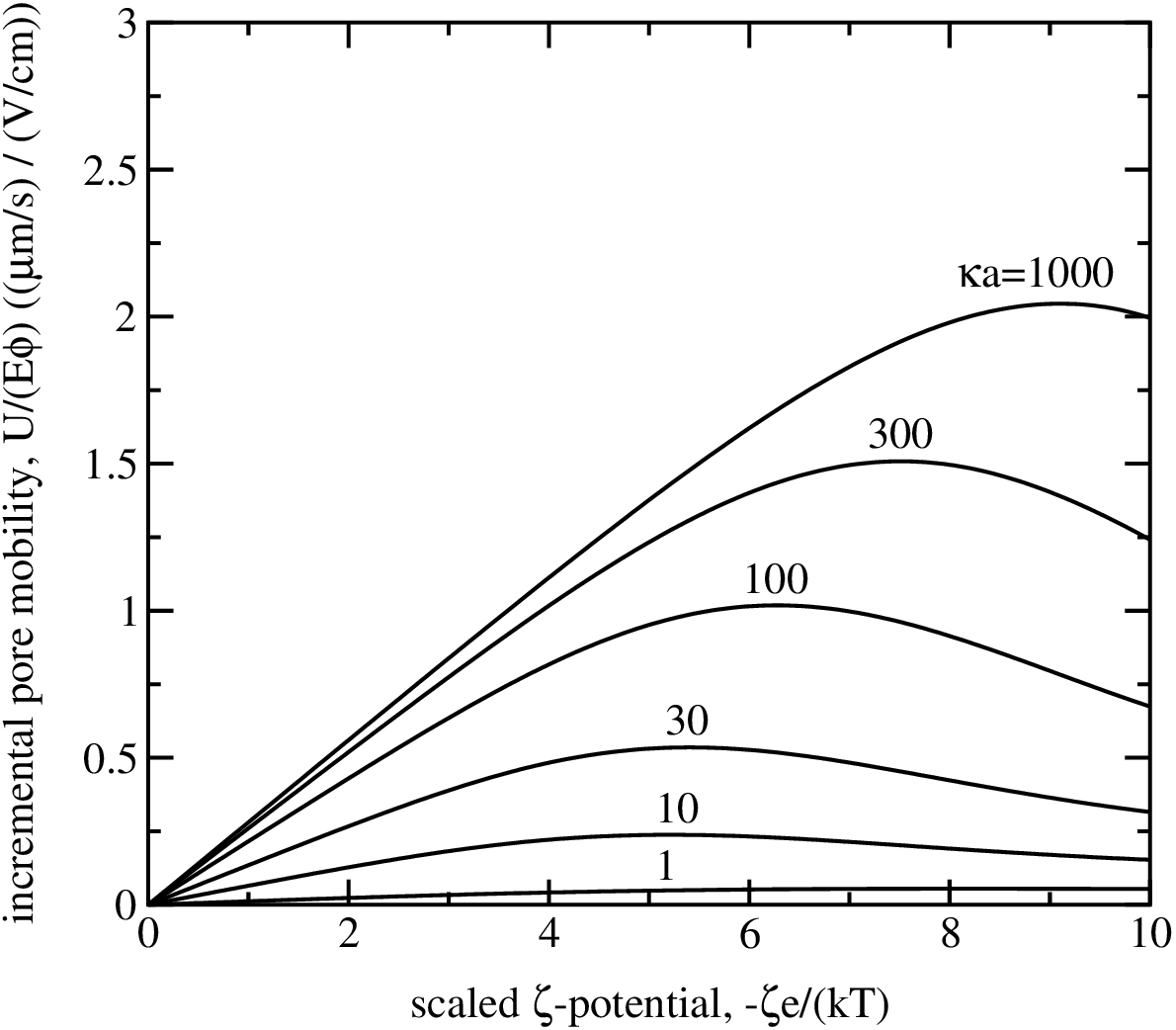}
    \hspace{0.75cm} \includegraphics[width=6cm]{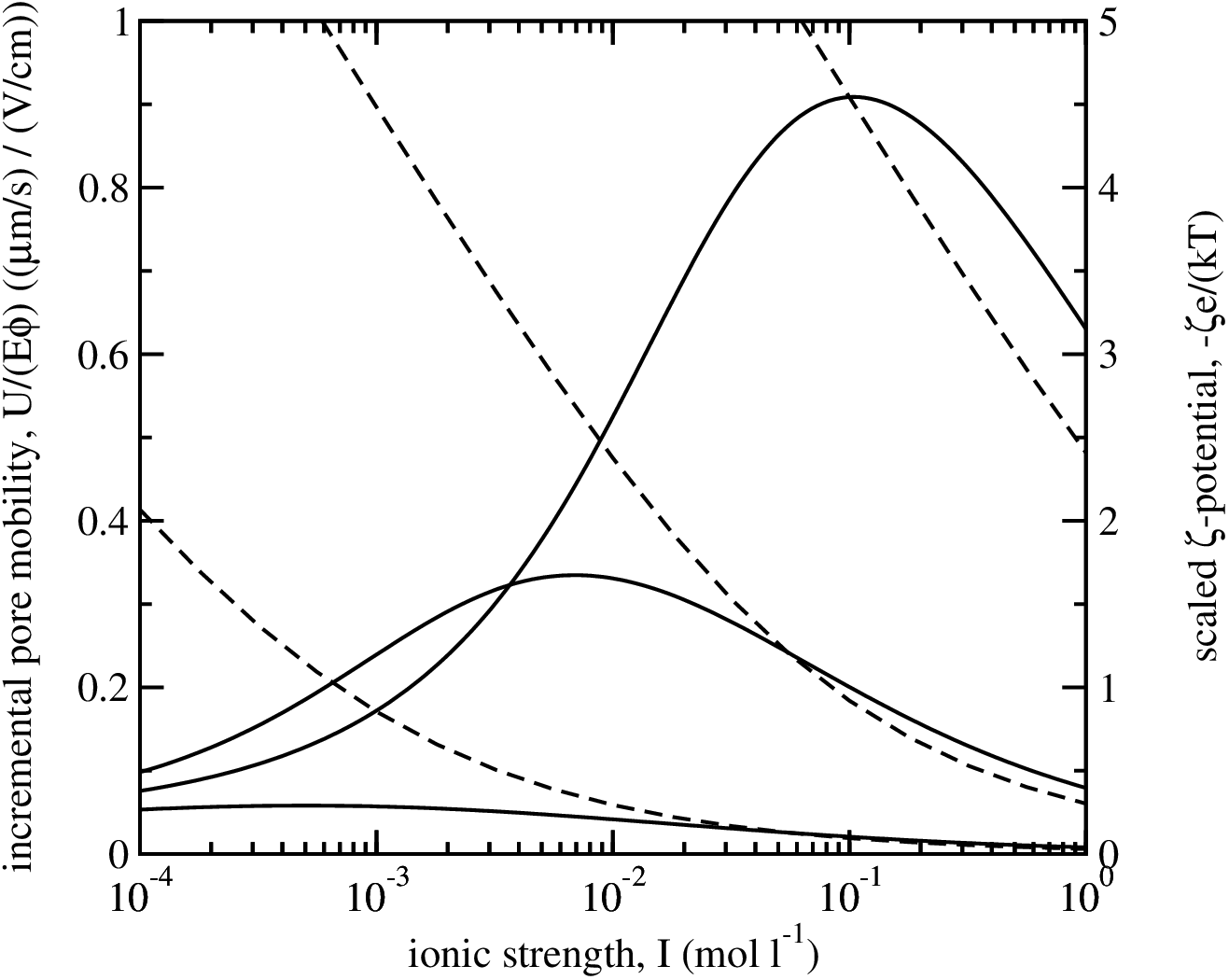}\\
    \vspace{1.0cm} \includegraphics[width=5.5cm]{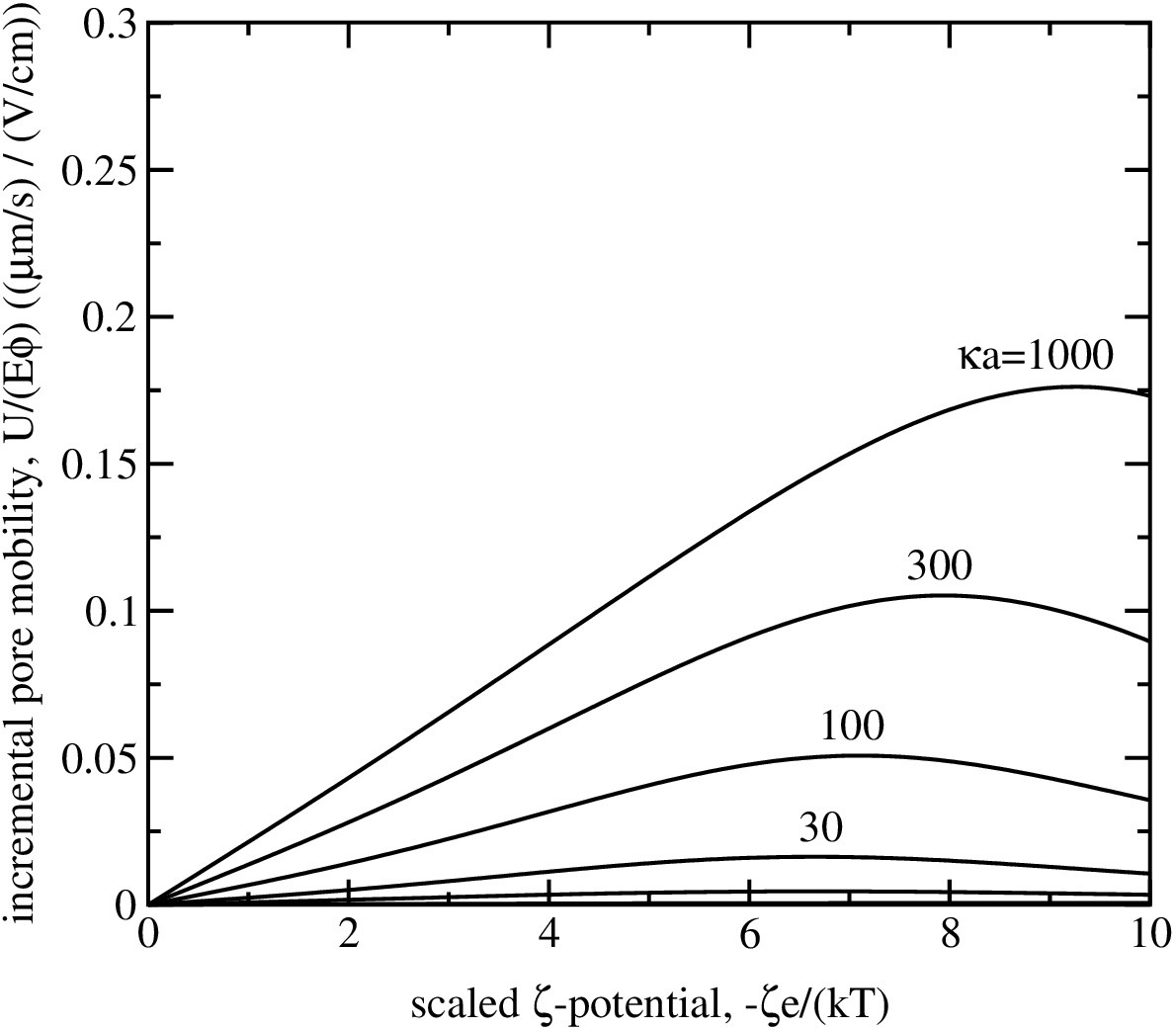}
    \hspace{0.75cm} \includegraphics[width=6cm]{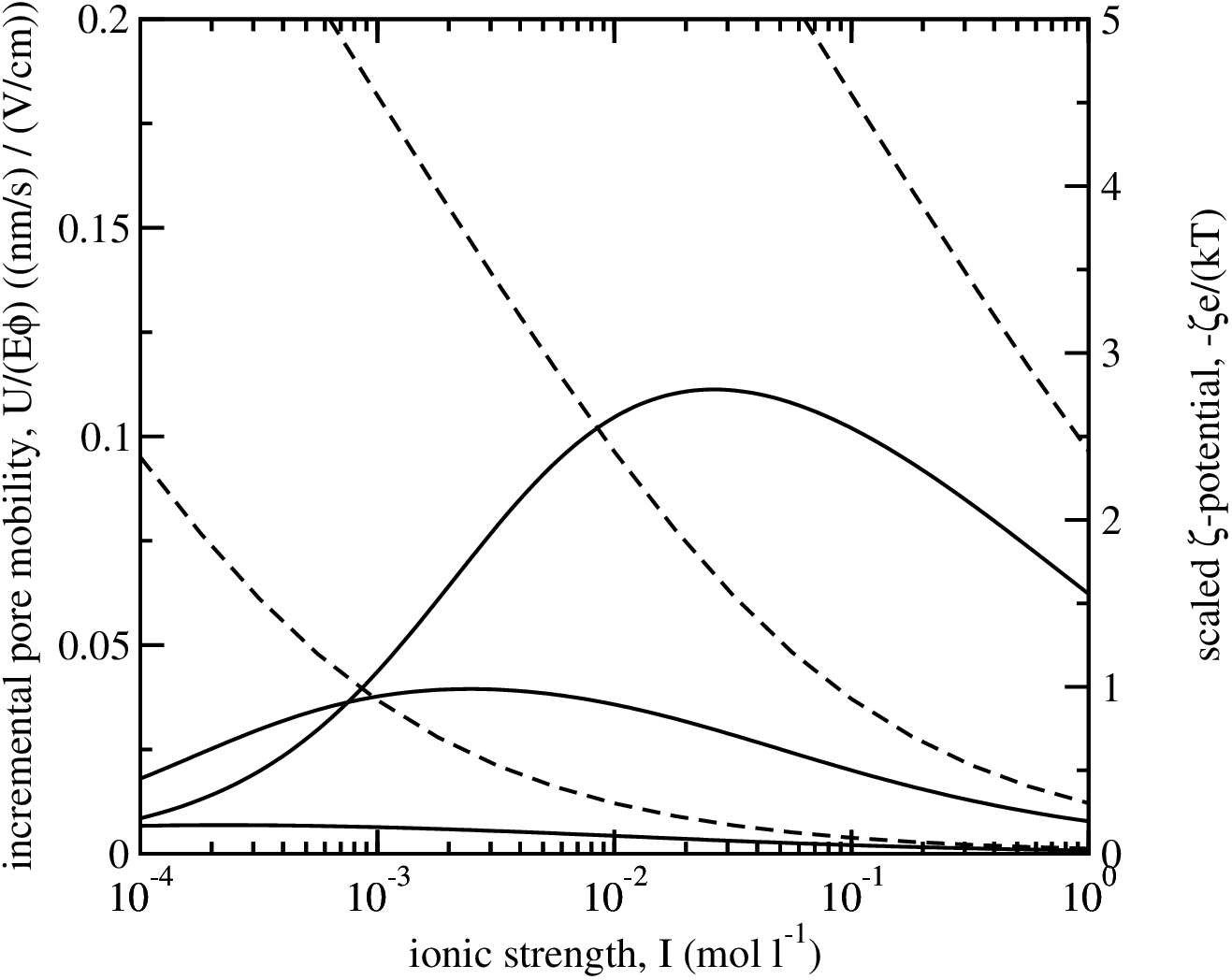}
    \vspace{0.5cm}
  \end{center}
  \caption{\label{fig:mobilityEBSL=3.12NM} The incremental pore
    mobility $U / (E \phi)$ with inclusion radii $a = 10$ ({\em top}),
    100 ({\em middle}) and 1000~nm ({\em bottom}): aqueous NaCl at $T
    = 25$\degc; $\bsl \approx 3.11$~nm. {\em Left panels} show the
    mobility as a function of the (scaled) $\zeta$-potential $\zeta
    e/(\kb T)$ for various (scaled) reciprocal double-layer
    thicknesses $\ka = 1$, 10, 30, 100, 300 and 1000. {\em Right
    panels} show the mobility ({\em solid lines}, left axis) and
    $\zeta$-potential ({\em dashed lines}, right axis) as a function
    of the ionic strength with constant surface charge densities $-
    \sigma \approx 0.179$, 1.79 and 17.9~$\mu$C~cm$^{-2}$ (increasing
    upward at high ionic strength).}
\end{figure}

The pore mobility is shown in the {\em left panels} of
figure~\ref{fig:mobilityEBSL=3.12NM} as a function of the
$\zeta$-potential for various values of $\ka$, with inclusion radii $a
= 10$, 100 and 1000~nm ({\em top-to-bottom panels}), and Brinkman
screening length $\bsl \approx 3.11$~nm. Similarly to the
electrophoretic mobility~\citep[\eg,][]{OBrien:1978}, at low to
moderate $\zeta$-potentials the pore mobility provides a one-to-one
connection between the (measured) pore velocity and the surface
charge. Mobility maxima arise from polarization (by electromigration)
and relaxation (by diffusion) of the equilibrium double layer.  As
suggested by earlier theoretical studies examining the role of
polarization and relaxation on the electrophoretic mobility of
polymer-coated particles~\citep{Saville:2000,Hill:2004a,Hill:2005a},
these calculations clearly demonstrate that polarization is driven by
electromigration, since convection is extremely weak when the
particles are immobilized in a polymer gel. Note that theoretical
studies of electroosmotic flow in micro-porous membranes do not reveal
such maxima, because the underlying microscale model comprises
(effectively) straight channels with charged walls~\citep{Yao:2003a}.

As expected, the (incremental) pore mobility tends to increase with
$\zeta$-potential at fixed ionic strength, and increases with ionic
strength at fixed $\zeta$-potential. Both trends reflect the
increasing charge required to maintain a constant surface potential
when varying the ionic strength. For colloids whose surface charge is
independent of electrolyte concentration, the $\zeta$-potential
increases with decreasing $\ka$. Accurate semi-empirical expressions
for this relationship (obtained from solutions of the
Poisson-Boltzmann equation) are readily
available~\citep{Russel:1989}. In general, however, the dependence of
surface charge density on ionic strength and pH is exceedingly
difficult to predict, and must therefore be determined empirically for
specific interfaces~\citep[\eg, see][for silica in the presence of
KCl]{Yao:2003b}.

To draw a closer connection to experiments, the pore mobility is shown
in the {\em right panels} of figure~\ref{fig:mobilityEBSL=3.12NM} as a
function of the ionic strength with three {\em constant} surface
charge densities spanning two orders of magnitude. Note that the
inclusion radii are the same as in the corresponding {\em left
panels}. Because the surface charge is fixed, the $\zeta$-potential
({\em dashed lines}, right axis) decreases with increasing ionic
strength, but the particle size does not significantly influence the
$\zeta$-potential. Because the average velocity reflects the combined
influence of all particles in the composite, $U$ is expected to be
proportional to the $O(n \sigma a^2 \sim \sigma \phi / a)$ (average)
counter-charge density. Therefore, balancing the corresponding $O(E
\sigma \phi / a)$ electrical force with the $O(\eta U / \bsl^2)$ Darcy
drag force gives a pore mobility $U / (\phi E) \sim \sigma \bsl^2 / (
\eta a)$. Indeed, comparing the mobility axes (left sides) of the
panels on the right-hand side of figure~\ref{fig:mobilityEBSL=3.12NM}
indicates that the mobility is, at least approximately, inversely
proportional to the inclusion radius. Again, at low ionic strengths,
when the $\zeta$-potential is high, polarization and relaxation
significantly complicate this simple interpretation.

\begin{figure}
  \begin{center}
    \vspace{1.0cm} \includegraphics[width=6cm]{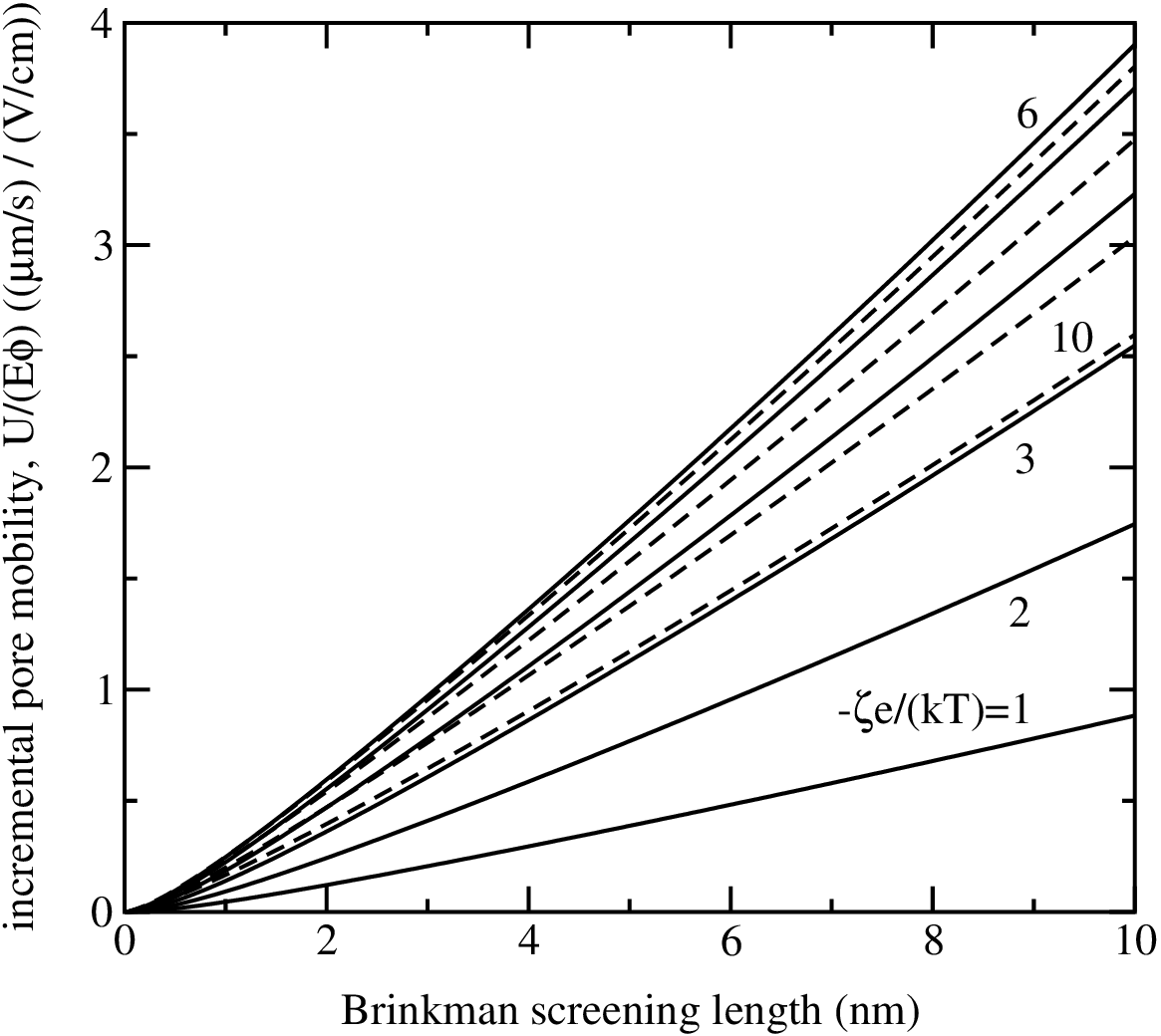}
    \vspace{0.5cm}
  \end{center}
  \caption{\label{fig:mobilityEka=100} The incremental pore mobility
    $U / (E \phi)$ as a function of the Brinkman screening length
    $\bsl$ for various (scaled) $\zeta$-potentials $-\zeta e/(\kb T) =
    1$, $2$, $3$, ..., $6$ ({\em solid lines}) $7$, ..., $10$ ({\em
    dashed lines}): aqueous NaCl at $T = 25$\degc; $a = 100$~nm; $\ka
    = 100$ ($I \approx 0.0925$~\M). An electric field is applied in
    the absence of average pressure and concentration gradients. The
    maximum velocity is achieved when $|\zeta| e / (\kb T) \approx
    6$.}
\end{figure}

\begin{figure}
  \begin{center}
    \vspace{1.0cm} \includegraphics[width=6cm]{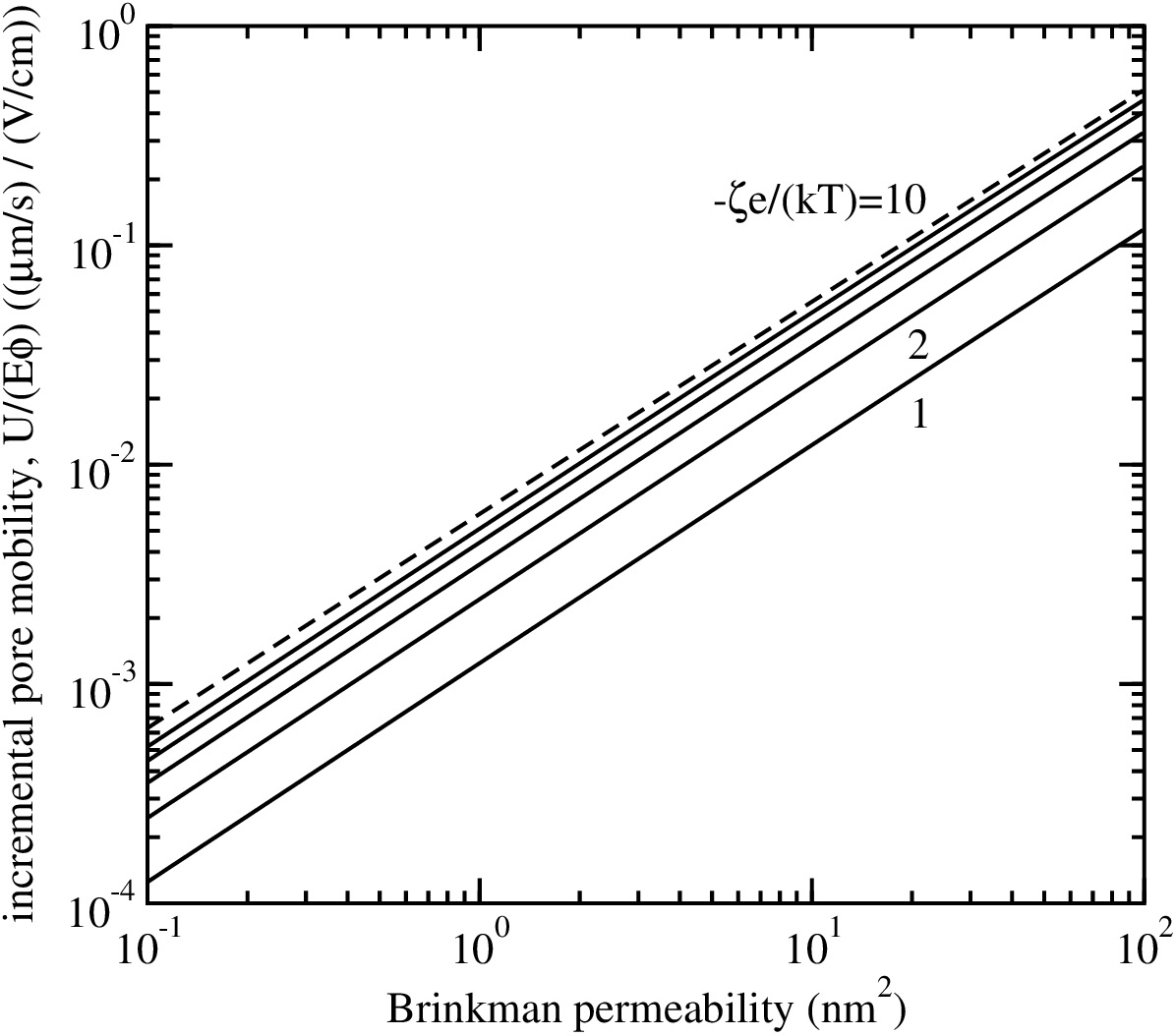}
    \vspace{0.5cm}
  \end{center}
  \caption{\label{fig:mobilityEka=1} The incremental pore mobility $U
    / (E \phi)$ as a function of the Darcy permeability $\bsl^2$ for
    various (scaled) $\zeta$-potentials $-\zeta e/(\kb T) = 1$, $2$,
    $3$, ..., $6$ ({\em solid lines}) $10$ ({\em dashed line}):
    aqueous NaCl at $T = 25$\degc; $a = 100$~nm; $\ka = 1$ ($I \approx
    9.25 \times 10^{-6}$~\M).}
\end{figure}

To highlight the influence of polymer-gel permeability, the pore
mobility is shown in figure~\ref{fig:mobilityEka=100} as a function of
the Brinkman screening length $\bsl$ for various
$\zeta$-potentials. With a particle radius $a = 100$~nm, the ionic
strength $I \approx 0.0925$~\M \ yields $\ka = 100$ and $\kappa \bsl >
1$ at most values of $\bsl$. Now the mobility increases as $\bsl^m$
with exponent $m(\bsl)$ in the range 1--2, indicating that viscous
stresses {\em and} Darcy drag balance the electrical body force. Note
that the mobility increases linearly with the $\zeta$-potential when
$|\zeta|$ is small, and, again, mobility maxima are evident when
$|\zeta| \approx 6 \kb T / e$. Pore mobilities are shown in
figure~\ref{fig:mobilityEka=1} at a much lower ionic strength $I
\approx 9.25 \times 10^{-6}$~\M \ yielding $\ka = 1$. Now, as
expected, the mobility increases linearly with $\bsl^2$, since $\kappa
\bsl < 1$. The monotonic increase with $\zeta$-potential is because
the surface charge densities are low and, hence, polarization is weak.

\subsection{Electroosmotic pumping} \label{sec:incporevel2}

Recall, the (incremental) pore mobility was defined with zero average
pressure gradient, and therefore neglects the pressure differential
$\Delta p = P L$ required to pump fluid through an external
network. This section briefly addresses coupling the composite and
electrodes---which together are referred to as an electroosmotic
pump---to a microfluidic network. The analysis briefly considers the
force exerted on the composite, electrical power consumption, and pump
efficiency.

Let us consider a closed loop, where fluid in the composite exits from
one side, passes through a microfluidic network, and returns to the
other side. To ensure that all electrical current flows through the
composite, we assume that the electrical resistance of the network is
much greater than that of the pump, which is realized when the length
(area) of the external network $L_e$ ($d^2$) is much greater (smaller)
than that of the composite $L$ ($A$). Next, assuming laminar viscous
flow, the pressure-drop through the network may be written
\begin{equation}
  \Delta p \approx \eta c (L_e / d^4) Q,
\end{equation}
where $d^2$ is the (characteristic) channel cross-sectional area, $Q$
is the volumetric flow rate, and $c$ is an $O(1)$ constant that
reflects the shape and length the network sections.

Equating the network pressure drop to the pump characteristic emerging
from Eqn.~(\ref{eqn:avemomB}),
\begin{eqnarray}
  \Delta p / L = - (\eta / \bsl^2) U - \phi (3 / a^3) (\eta / \bsl^2)
  (E C^E + U C^U),
\end{eqnarray}
gives
\begin{eqnarray} \label{eqn:pressuredrop}
  [\eta c (L_e / d^4) (A / L) + (\eta / \bsl^2) + \phi (3 / a^3) (\eta
    / \bsl^2) C^U ] Q / A = - \phi (3 / a^3) (\eta / \bsl^2) E C^E,
\end{eqnarray}
where $A$ and $L$ are the composite cross-sectional area and length.

The contribution of the back-flow, as represented by the asymptotic
coefficient $C^U$ in Eqn.~(\ref{eqn:pressuredrop}), is obtained from
the (U) problem. This and other asymptotic coefficients are provided
in table~\ref{tab:ufieldalonecoeffs} for a composite with Brinkman
screening length $\bsl \approx 0.951$~nm and inclusion radius $a =
100$~nm; again, the $\zeta$-potentials and (three) ionic strengths
span experimentally accessible ranges. Here, the average force exerted
by the polymer on the inclusions $\langle \vect{f}^m \rangle = - 4 \pi
(\eta / \bsl^2) C^U \vect{U}$ (Eqn.~(\ref{eqn:force3})) is independent
of the surface charge. Note that $C^U = (1/2) a^3$ when the only
contribution to the force is due to Darcy flow ($\bsl / a \ll 1$); the
(constant) value $C^U / a^3 \approx 0.514$ in
table~\ref{tab:ufieldalonecoeffs}, which reflects a small viscous
contribution ($\bsl / a = 0.00951$), is precisely the value given by
Brinkman's theory (Eqn.~(\ref{eqn:brinkman})). The flow-induced
electrical and concentration polarization, as represented by $D^U$ and
$C_j^U$, are related to the streaming potential and streaming current,
and are included here only for future reference.

   \begin{table}
  \begin{center}
    \caption{\label{tab:ufieldalonecoeffs} Dimensionless asymptotic
      coefficients for bulk {\em convection} of NaCl in a Brinkman
      medium with charged spherical inclusions: $a = 100$~nm;
      $\bsl\approx 0.951$~nm; $T = 25$\degc, $D_1 \approx 1.33 \times
      10^{-9}$m$^2$s$^{-1}$~(Na$^+$); $D_2 \approx 2.03 \times
      10^{-9}$m$^2$s$^{-1}$~(Cl$^-$); $u^* \approx 5.15 \times
      10^{-3}$~m~s$^{-1}$.}
    
    \begin{tabular*}{\columnwidth}{@{\extracolsep{\fill}}llll} \hline
      
      \multicolumn{1}{c}{$\zeta e /(\kb T)$} & \multicolumn{1}{c}{$D^U
	e u^* / (\kb T a^2)$} & \multicolumn{1}{c}{$C_j^U u^* / (2 I
	a^2)$} & \multicolumn{1}{c}{$C^U / a^3$} \\ & &
      \multicolumn{1}{c}{($j=1,2$)} & \\\hline
      
      $\ka=  1$ & $I=9.25\times 10^{-6}$~\M \\ \hline
      
      $-1$ & $6.70\times 10^{-1}$ & $8.71\times 10^{-2}$ & $5.14\times 10^{-1}$ \\
      $-2$ & $1.32\times 10^{0}$  & $2.03\times 10^{-1}$ & $5.14\times 10^{-1}$ \\
      $-4$ & $2.44\times 10^{0}$  & $4.49\times 10^{-1}$ & $5.14\times 10^{-1}$ \\
      $-6$ & $3.10\times 10^{0}$  & $6.06\times 10^{-1}$ & $5.14\times 10^{-1}$ \\
      $-8$ & $3.33\times 10^{0}$  & $6.44\times 10^{-1}$ & $5.14\times 10^{-1}$ \\ \hline
      
      $\ka= 10$ & $I=9.25\times 10^{-4}$~\M \\ \hline
      
      $-1$ & $4.38\times 10^{-2}$ & $7.76\times 10^{-4}$ & $5.14\times 10^{-1}$ \\
      $-2$ & $9.10\times 10^{-2}$ & $2.16\times 10^{-3}$ & $5.14\times 10^{-1}$ \\
      $-4$ & $1.81\times 10^{-1}$ & $5.58\times 10^{-3}$ & $5.14\times 10^{-1}$ \\
      $-6$ & $2.19\times 10^{-1}$ & $7.08\times 10^{-3}$ & $5.14\times 10^{-1}$ \\
      $-8$ & $1.96\times 10^{-1}$ & $5.74\times 10^{-3}$ & $5.14\times 10^{-1}$ \\ \hline
      
      $\ka=100$ & $I=9.25\times 10^{-2}$~\M \\ \hline
      
      $-1$ & $2.55\times 10^{-3}$ & $4.49\times 10^{-6}$ & $5.14\times 10^{-1}$ \\
      $-2$ & $5.33\times 10^{-3}$ & $1.27\times 10^{-5}$ & $5.14\times 10^{-1}$ \\
      $-4$ & $1.14\times 10^{-2}$ & $3.69\times 10^{-5}$ & $5.14\times 10^{-1}$ \\
      $-6$ & $1.64\times 10^{-2}$ & $5.93\times 10^{-5}$ & $5.14\times 10^{-1}$ \\
      $-8$ & $1.70\times 10^{-2}$ & $6.14\times 10^{-5}$ & $5.14\times 10^{-1}$ \\
      
    \end{tabular*}
  \end{center}
  \end{table}

When the dominant resistance comes from the composite itself (the
second term on the left-hand-side of Eqn.~(\ref{eqn:pressuredrop})),
the pump performance curve simplifies to
\begin{eqnarray}
  Q \approx - \phi (3 / a^3) A E C^E,
\end{eqnarray}
which is clearly independent of the applied load. Neglecting
constraints imposed by electrolysis, for example, the maximum length
of the composite may be set by consideration of the electrical power
consumption
\begin{equation}
  {\cal P} \approx K^\infty E^2 A L + O(\phi).
\end{equation}

Note that the force exerted on the composite
\begin{equation} \label{eqn:compforce}
  F = -\Delta p A \approx \eta c (L_e / d^4) A^2 \phi (3 / a^3) E C^E
\end{equation}
reflects the pressure required to pump fluid through the
network. Therefore, the average shear stress $\tau \sim - F / (L
\sqrt{A})$ required to support the composite scales as
\begin{equation} \label{eqn:stress}
  \tau \sim - \eta c (L_e / d^4) (A^{3/2} / L) \phi (3 / a^3) E C^E.
\end{equation}
Since the area $A$ will be set by the flow rate, the length of the
composite should be set by the maximum allowable shear stress.

Finally, the pump efficiency, as measured by the ratio of the rate of
flow work $|Q \Delta p |$ to the electrical power consumption ${\cal
P}$ is
\begin{equation}
  {\cal E} \approx \eta c (L_e / d^4) [\phi (3 / a^3) C^E]^2 A /
  (K^\infty L).
\end{equation}
Because $C^E$ depends on $\zeta$, $\ka$, $\bsl$, etc., care must be
taken in interpreting this equation. Nevertheless, geometrical
considerations alone clearly favor thin membranes with large
cross-sectional area, operating with a low ionic strength
(conductivity) and a high inclusion volume fraction. In practice, an
optimal design (with specified flow $Q$ and, perhaps, voltage $V =
\Delta \psi$) will be constrained by consideration of the mechanical
strength of the composite (as indicated by Eqns.~(\ref{eqn:compforce})
and (\ref{eqn:stress})), which clearly diminishes with decreasing
thickness $L$ and increasing area $A$.

\subsection{Incremental pressure gradient} \label{sec:incporevel3}

Now consider the pressure gradient produced by an average electric
field with zero average flow. This situation may be realized when an
electrolyte-saturated composite is bounded by a vessel with
impenetrable walls. A practical application involves measuring the
differential (static) pressure $\Delta p$ to infer the permeability of
the polymer gel or, for example, the $\zeta$-potential of the
inclusions. Note that zero average flow does not imply stationary
fluid at the microscale, because the `inner' electroosmotic flow
around each inclusion is balanced by a far-field pressure-driven
back-flow, analogous to the situation encountered in
microelectrophoresis capillaries with blocked ends.

Setting $U = 0$ in Eqn.~(\ref{eqn:avemomC}) gives
\begin{equation}
  \vect{P} = -\phi (\eta / \bsl^2) (3 / a^3) C^E \vect{E} + O(\phi^2),
\end{equation}
which is termed the {\em incremental pressure
gradient}. Representative values may be calculated by multiplying the
incremental pore mobilities in the last column of
table~\ref{tab:efieldalonecoeffs}, and plotted in
figures~\ref{fig:mobilityEBSL=3.12NM}--\ref{fig:mobilityEka=1}, by
their respective values of $\eta / \bsl^2$.

It is interesting to note that when $\kappa \bsl < 1$ and, hence, the
pore mobility increases linearly with $\bsl^2$ (see
figure~\ref{fig:mobilityEka=1}), the pressure gradient is independent
of the permeability. This is because increasing the permeability
increases the electric-field-induced flow, and, since the back-flow
and accompanying pressure gradient are proportional to each other, it
follows that the pressure gradient is independent of
$\bsl^2$. However, at higher ionic strengths, when $\kappa \bsl > 1$
(see figure~\ref{fig:mobilityEka=100}), the pressure gradient
evidently decreases with $\bsl^2$. This is because the
electric-field-induced flow within the diffuse double layers---where
resistance to flow is predominantly due to viscous stress---increases
more slowly with the permeability than the Darcy drag beyond the
double layers decreases.

Note that the incremental pressure gradient reflects the same
asymptotic coefficient $C^E$ as the pore mobility, so it is important
to establish whether measuring the electric-field-induced pressure
gradient offers a significant advantage over measuring the pore
mobility. Recall, pore mobilities can generate low but measurable
velocities in a microchannel. From table~\ref{tab:efieldalonecoeffs}
with $\ka = 100$ and $\zeta = - 1 \kb T / e$, setting $E =
2$~V~cm$^{-1}$ and $\phi = 10^{-2}$ yields $P \approx
8.76$~kPa~cm$^{-1}$. Therefore, when $L = 5$~mm, for example, the
pressure differential is $|\Delta p| \approx 4.38$~kPa (static head of
0.43~m of water). Clearly, the pressure gradient induced by a
relatively weak electric field is sufficient to produce a modest
(static) pressure.

\subsection{Species fluxes} \label{sec:specfluxes}

Let us write the average flux of each species from
Eqn.~(\ref{eqn:fluxE}) as
\begin{equation} \label{eqn:migrationcoeffs}
  \vect{J}_j = z_j e \frac{D_j}{\kb T} n_j^\infty \vect{E} (1 + \phi
  \Delta_j^E),
\end{equation}
where
\begin{eqnarray} \label{eqn:migrationinc}
  \Delta_j^E &=& \Delta_{j,e}^E + \Delta_{j,d}^E + \Delta_{j,c}^E
  \nonumber \\ &=& (3/a^3) D^E + (3 / a^3) \frac{\kb T}{z_j e
    n_j^\infty} C_j^E + (3/ a^3) \frac{\kb T}{z_j e D_j} C^E
\end{eqnarray}
is the sum of incremental {\em microscale} contributions (of
electromigration, diffusion, and convection, respectively) to the
average electromigrative flux.

The ratio of the convective and electromigrative terms is
\begin{equation}
  \Delta_{j,c}^E / \Delta_{j,e}^E = \frac{C^E}{z_j e \frac{D_j}{\kb T}
    D^E} \sim \Pe_j / z_j,
\end{equation}
and the ratio of the convective and diffusive terms is
\begin{equation}
  \Delta_{j,c}^E / \Delta_{j,d}^E = \frac{n_j^\infty C^E}{D_j C_j^E}
  \sim \Pe_j,
\end{equation}
where the P{\'e}clet number $\Pe_j = u_c \kappa^{-1} / D_j$ is
typically very small. The characteristic (microscale) velocity $u_c$
may be estimated by balancing the $O(E \sigma \kappa)$ electrical
force (per unit volume) with the $O(u_c \eta / \bsl^2)$ Darcy drag
force, giving $u_c \sim E \sigma \kappa \bsl^2 / \eta$. With $E \sim
10^2$~V~cm$^{-1}$, $\sigma \sim 1$~$\mu$C~cm$^{-2}$, $\kappa^{-1} \sim
10^2$~nm, $\bsl \sim 1$~nm, and $\eta \sim
10^{-3}$~kg~m$^{-1}$s$^{-1}$, $u_c \sim 1$~$\mu$m~s$^{-1}$. Further,
with $a \sim 1$~$\mu$m and $D_j \sim 10^{-9}$~m$^2$s$^{-1}$, $\Pe_j
\sim 10^{-4}$, indicating that diffusion and electromigration ($z_j
\ne 0$) dominate convection. Clearly, for charged species with $z_j
\sim 1$, electromigrative fluxes are comparable to diffusive fluxes.

Incremental contributions to the ion fluxes are provided in
table~\ref{tab:efieldaloneincs} for the composite whose asymptotic
coefficients are shown in table~\ref{tab:efieldalonecoeffs}. These
confirm that the convective contribution $\Delta_{j,c}^E$ is small,
and that the electromigrative contribution $\Delta_{j,e}^E$ approaches
the Maxwell value $\Delta^E_{j,e} = -3/2$ for impenetrable
(non-conducting) spheres as $|\zeta| \rightarrow 0$. Furthermore, the
electromigrative ($\Delta_{j,e}^E$) and diffusive ($\Delta_{j,d}^E$)
contributions for Na$^+$ and Cl$^-$ vary significantly with ionic
strength. For example, at low electrolyte concentration, the diffusive
term dominates, enhancing the flux of the counter-ion (Na$^+$) and
attenuating that of the co-ion (Cl$^-$).

  \begin{sidewaystable}
    \small 
    \begin{center}
      \caption{\label{tab:efieldaloneincs} Incremental contributions
	(see Eqn.~(\ref{eqn:migrationinc})) to the bulk {\em
	electromigration} of NaCl in a Brinkman medium with charged
	spherical inclusions: $a = 100$~nm; $\bsl \approx 0.951$~nm;
	$T = 25$\degc; $D_1 \approx 1.33 \times
	10^{-9}$m$^2$s$^{-1}$~(Na$^+$); $D_2 \approx 2.03 \times
	10^{-9}$m$^2$s$^{-1}$~(Cl$^-$).}
      
      \begin{tabular*}{\columnwidth}{@{\extracolsep{\fill}}llllllll} \hline
	 \multicolumn{1}{c}{$\zeta e /(\kb T)$} &
	 \multicolumn{1}{c}{$\Delta_{j,e}^E$ ($j=1,2$)} &
	 \multicolumn{1}{c}{$\Delta_{1,d}^E$~(Na$^+$) ($=-\Delta_{2,d}^E$)} &
	 \multicolumn{1}{c}{$\Delta_{1,c}^E$~(Na$^+$)} &
	 \multicolumn{1}{c}{$\Delta_{2,c}^E$~(Cl$^-$)} &
	 \multicolumn{1}{c}{$\Delta_1^E$~(Na$^+$)} &
	 \multicolumn{1}{c}{$\Delta_2^E$~(Cl$^-$)} &
	 \multicolumn{1}{c}{$\Delta^K$}\\ \hline
	
	$\ka=  1$ & $I=9.25\times 10^{-6}$~\M \\ \hline
	$-1$ & $-1.15\times 10^{0}$  & $ 6.25\times 10^{0}$ & $-2.17\times 10^{-4}$ & $1.42\times 10^{-4}$ & $5.10\times 10^{0}$ & $-7.39\times 10^{0}$ & $-2.44\times 10^{0}$\\
	$-2$ & $-1.71\times 10^{-1}$ & $ 1.22\times 10^{1}$ & $-4.25\times 10^{-4}$ & $2.78\times 10^{-4}$ & $1.21\times 10^{1}$ & $-1.24\times 10^{1}$ & $-2.72\times 10^{0}$\\
	$-4$ & $+2.60\times 10^{0}$  & $ 2.23\times 10^{1}$ & $-7.70\times 10^{-4}$ & $5.05\times 10^{-4}$ & $2.49\times 10^{1}$ & $-1.97\times 10^{1}$ & $-2.05\times 10^{0}$\\
	$-6$ & $+4.79\times 10^{0}$  & $ 2.86\times 10^{1}$ & $-9.74\times 10^{-3}$ & $6.38\times 10^{-4}$ & $3.34\times 10^{1}$ & $-2.39\times 10^{1}$ & $-1.16\times 10^{0}$\\
	$-8$ & $+5.89\times 10^{0}$  & $ 3.17\times 10^{1}$ & $-1.05\times 10^{-3}$ & $6.87\times 10^{-4}$ & $3.76\times 10^{1}$ & $-2.58\times 10^{1}$ & $-6.91\times 10^{-1}$\\ \hline
	
	$\ka= 10$ & $I=9.25\times 10^{-4}$~\M \\ \hline
	$-1$ & $-1.42\times 10^{0}$  & $ 4.38\times 10^{-1}$ & $-1.39\times 10^{-3}$ & $9.12\times 10^{-4}$ & $-9.85\times 10^{-1}$ & $-1.86\times 10^{0}$ & $-1.51\times 10^{0}$\\
	$-2$ & $-1.19\times 10^{0}$  & $ 9.08\times 10^{-1}$ & $-2.81\times 10^{-3}$ & $1.84\times 10^{-3}$ & $-2.88\times 10^{-1}$ & $-2.10\times 10^{0}$ & $-1.38\times 10^{0}$\\
	$-4$ & $-4.11\times 10^{-1}$ & $ 1.95\times 10^{0}$  & $-5.36\times 10^{-3}$ & $3.52\times 10^{-3}$ & $+1.53\times 10^{0}$  & $-2.35\times 10^{0}$ & $-8.16\times 10^{-1}$\\
	$-6$ & $+3.96\times 10^{-1}$ & $ 2.86\times 10^{0}$  & $-6.32\times 10^{-3}$ & $4.14\times 10^{-3}$ & $+3.25\times 10^{0}$  & $-2.46\times 10^{0}$ & $-1.98\times 10^{-1}$\\
	$-8$ & $+8.84\times 10^{-1}$ & $ 3.38\times 10^{0}$ &  $-5.45\times 10^{-3}$ & $3.57\times 10^{-3}$ & $+4.26\times 10^{0}$  & $-2.50\times 10^{0}$ & $+1.81\times 10^{-1}$\\ \hline
	
	$\ka=100$ & $I=9.25\times 10^{-2}$~\M \\ \hline
	$-1$ & $-1.49\times 10^{0}$  & $ 4.70\times 10^{-2}$ & $-8.08\times 10^{-3}$ & $5.30\times 10^{-3}$ & $-1.45\times 10^{0}$  & $-1.53\times 10^{0}$ & $-1.50\times 10^{0}$\\
	$-2$ & $-1.45\times 10^{0}$  & $ 1.08\times 10^{-1}$ & $-1.64\times 10^{-2}$ & $1.08\times 10^{-2}$ & $-1.36\times 10^{0}$  & $-1.54\times 10^{0}$ & $-1.47\times 10^{0}$\\
	$-4$ & $-1.23\times 10^{0}$  & $ 3.35\times 10^{-1}$ & $-3.33\times 10^{-2}$ & $2.18\times 10^{-2}$ & $-9.31\times 10^{-1}$ & $-1.55\times 10^{0}$ & $-1.30\times 10^{0}$\\
	$-6$ & $-7.59\times 10^{-1}$ & $ 8.11\times 10^{-1}$ & $-4.44\times 10^{-2}$ & $2.91\times 10^{-2}$ & $+7.38\times 10^{-3}$ & $-1.54\times 10^{0}$ & $-9.27\times 10^{-1}$\\
	$-8$ & $-1.14\times 10^{-1}$ & $ 1.46\times 10^{0}$  & $-4.15\times 10^{-2}$ & $2.72\times 10^{-2}$ & $+1.30\times 10^{0}$  & $-1.55\times 10^{0}$ & $-4.17\times 10^{-1}$\\
	
      \end{tabular*}
    \end{center}
  \end{sidewaystable}

\subsection{Electrical conductivity} \label{sec:eleccond}

The electrical conductivity of colloidal dispersions is well known to
reflect the particle surface charge density~\citep{Russel:1989}. The
conductivity of a composite with immobilized particles presents a
relatively simple problem when the electrolyte ions are unhindered by
the polymer, because only the electrolyte ions---not the charge on the
particles (macroions) themselves---contribute to charge transfer. This
section establishes whether the electrical conductivity is sensitive
to the surface charge and, possibly, the permeability of the polymer
gel.

From the fluxes in Eqn.~(\ref{eqn:fluxE}), the average current density
may be written
\begin{eqnarray} \label{eqn:current}
  \vect{I} = \sum_{j=1}^{N} z_j e \vect{J}_j \approx K^\infty \vect{E}
  + \phi (3 /a^3) \vect{E} [K^\infty D^E + \sum_{j=1}^{N} z_j e D_j
  C_j^E],
\end{eqnarray}
where
\begin{equation} \label{eqn:bulkcond}
  K^\infty = \sum_{j=1}^{N} (z_j e)^2 \frac{D_j}{\kb T} n_j^\infty
\end{equation}
is the conductivity of the electrolyte. The conductivity of the
composite is defined as
\begin{equation} \label{eqn:effcond}
  K^* = I / E = K^\infty (1 + \phi \Delta^K),
\end{equation}
where $\Delta^K$ is termed the (dimensionless) {\em conductivity
increment}.

Equations~(\ref{eqn:current})-(\ref{eqn:effcond}) are equivalent to
expressions derived by~\cite{OBrien:1981} for the conductivity of
dilute colloidal dispersions with the particles undergoing
electrophoresis. Here, the asymptotic coefficients are different
because the particles are stationary\footnote{When particles undergo
electrophoresis in a Newtonian electrolyte, the far-field (solenoidal)
velocity disturbance is irrotational, decaying as $r^{-3}$ as $r
\rightarrow \infty$. When {\em fixed} in electrolyte without polymer,
however, the far-field disturbance reflects a net force, decaying as
$r^{-1}$ as $r \rightarrow \infty$.}. When the $\zeta$-potential is
low and, hence, ion fluxes are unperturbed by the surface charge, the
dipole strength for non-conducting spheres equals the Maxwell value
$D^E = - (1/2) a^3$, so $\Delta^K \rightarrow -3/2$ as $|\zeta|
\rightarrow 0$. In general, however, the conductivity increment also
reflects the charge of the inclusions {\em and} the ionic strength.

Similarly to dispersions, the average convective term (involving $C^E$
in Eqn.~(\ref{eqn:fluxE})) does not influence the conductivity
increment (because of electrical neutrality), and the diffusive term
(involving $C_j^E$) vanishes only when the species have equal
mobilities. In general, the (microscale) electromigrative and
diffusive terms (involving $D^E$ and $C_j^E$) contribute to the
average current density. However, because these are influenced by
fluid motion, the conductivity of a composite is not the same as when
particles are fixed in an electrolyte without polymer. As expected,
because only the electrolyte ions contribute to charge transfer, the
conductivity of a composite is lower than when particles undergo
electrophoresis.

\begin{figure}
  \begin{center}
    \vspace{1.0cm}
    \includegraphics[width=6cm]{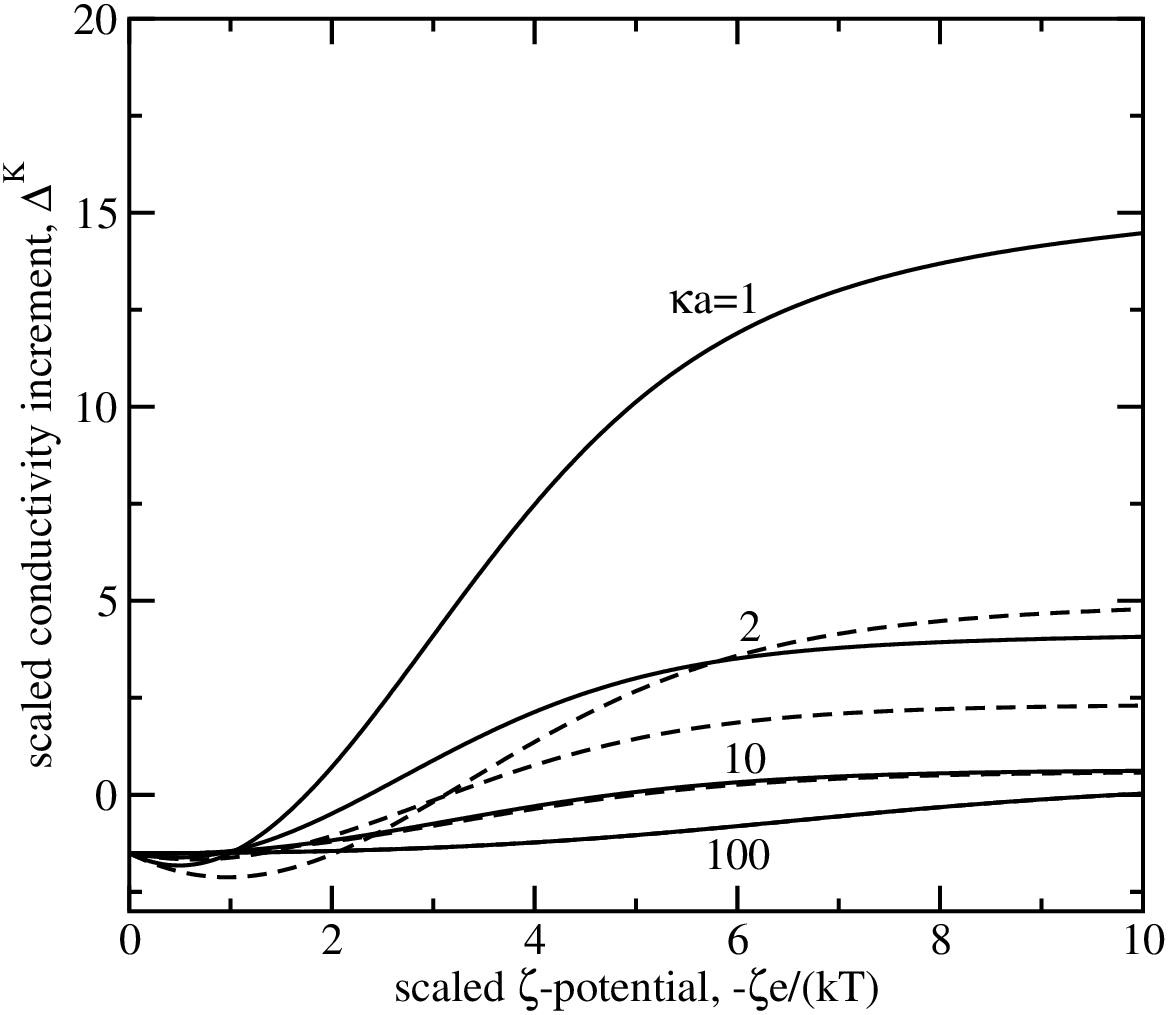}
    \vspace{0.5cm}
  \end{center}
  \caption{\label{fig:condincEfree} The (scaled) conductivity
    increment $\Delta^K$ (see Eqn.~(\ref{eqn:effcond})) as a function
    of the (scaled) $\zeta$-potential $\zeta e /(\kb T)$ for various
    (scaled) reciprocal double-layer thicknesses $\ka = 1$, 2, 10 and
    100 (aqueous NaCl at $T = 25$\degc \ with $a = 100$~nm) for
    particles undergoing electrophoresis ({\em solid lines}) and
    stationary particles ({\em dashed lines}), both in electrolyte
    without polymer.}
\end{figure}

Representative conductivity increments for mobile ({\em solid lines})
and stationary ({\em dashed lines}) particles in a NaCl electrolyte
without polymer are compared in
figure~\ref{fig:condincEfree}\footnote{The computations for particles
in a pure electrolyte were performed with software (called MPEK,
available from the author) based on the work
of~\cite{Hill:2003a}. These accurately reproduced earlier calculations
by~\cite{OBrien:1981} for KCl and HClO$_4$ electrolytes.}. At low
ionic strength (small $\ka$), immobilizing the particles decreases the
conductivity increment, because particle migration (electrophoresis)
contributes significantly to charge transfer. At high ionic strength
($\ka > 10$), however, the conductivity increments for mobile and
fixed particles are (practically) the same. This is because the
density of charge added by the particles (macroions) is vanishingly
small compared to the density of bulk electrolyte ions, so the
conductivity reflects only the contribution of (dielectric and
double-layer) polarization to the average electric field.

In contrast to KCl and HClO$_4$
electrolytes~\citep[see][]{OBrien:1981}, the conductivity increment
with NaCl {\em decreases} with increasing $\zeta$-potential when
$\zeta$ and $\ka$ are small. This may be attributed to the counter-ion
(Na$^+$) having a significantly lower mobility than the co-ion
(Cl$^-$). For KCl, the mobilities of K$^+$ and Cl$^-$ are very
similar, yielding a monotonically increasing conductivity
increment. For HClO$_4$, however, the counter-ion (H$^+$) has a
significantly higher mobility than the co-ion (ClO$_4^-$), yielding a
conductivity increment that {\em increases} more rapidly (linearly)
with $|\zeta|$~\citep[see][]{OBrien:1981}.

\begin{figure}
  \begin{center}
    \vspace{1.0cm} \includegraphics[width=6cm]{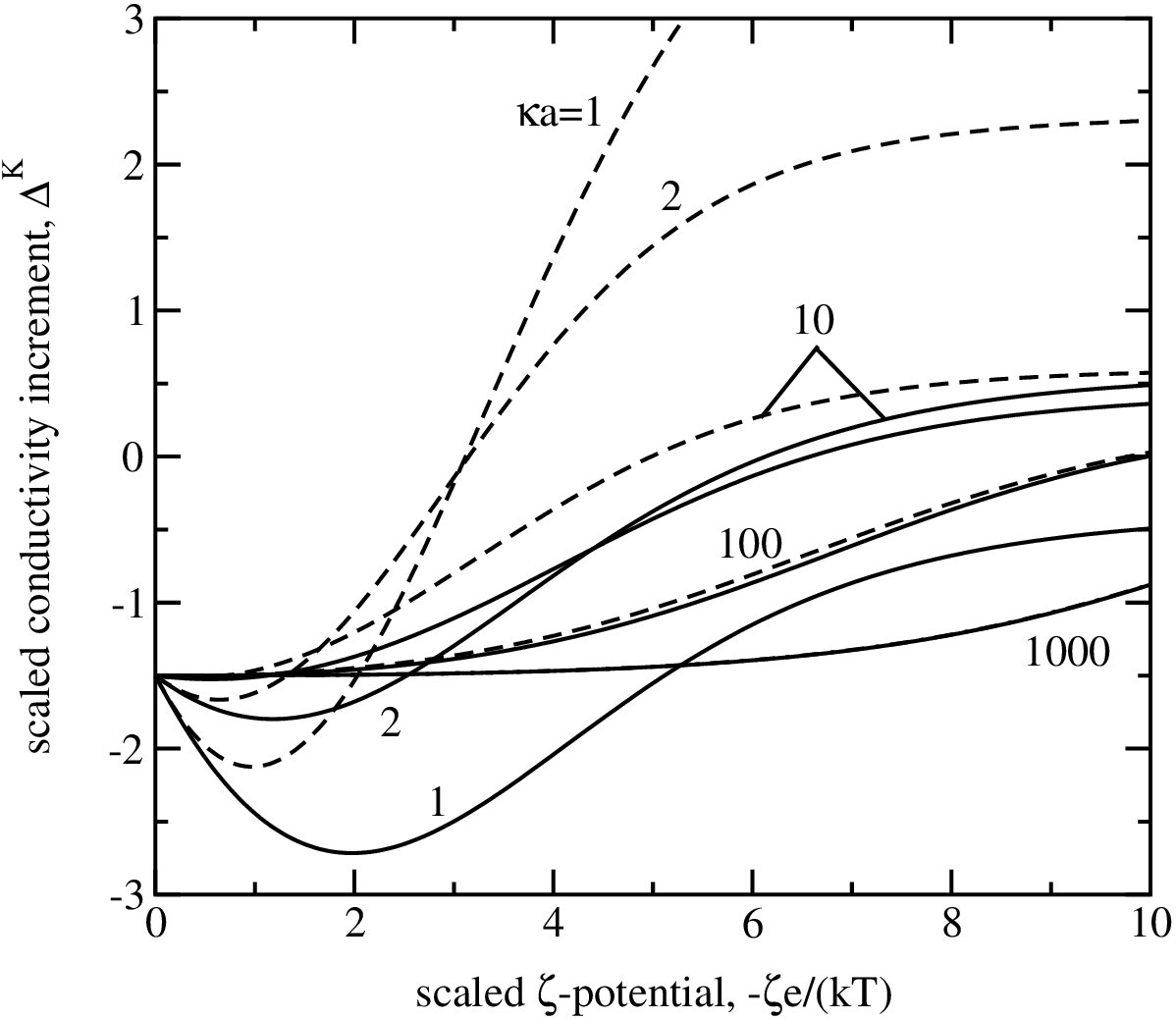}
    \vspace{0.5cm}
  \end{center}
  \caption{\label{fig:condincE} The (scaled) conductivity increment
    $\Delta^K$ (see Eqn.~(\ref{eqn:effcond})) as a function of the
    (scaled) $\zeta$-potential $\zeta e / (\kb T)$ for various
    (scaled) reciprocal double-layer thicknesses $\ka = 1$, 2, 10, 100
    and 1000: aqueous NaCl at $T = 25$\degc; $a = 100$~nm; $\bsl
    \approx 3.11$ ({\em solid lines}); stationary particles in
    electrolyte without polymer ({\em dashed lines}).}
\end{figure}

The conductivity increment for particles with radius $a=100$~nm
embedded in a polymer gel with Brinkman screening length $\bsl \approx
3.11$~nm ({\em solid lines}) is shown in
figure~\ref{fig:condincE}. Similar calculations (not shown) reveal
that an order-of-magnitude increase in the permeability $\bsl^2$
produces almost the same values at all $\zeta$-potentials and ionic
strengths (values of $\ka$). This confirms that the average current
density is dominated by electromigration and diffusion. The
conductivity increments ({\em solid lines}) are compared to values for
stationary particles in an electrolyte without polymer ({\em dashed
lines}). These results are almost the same at high ionic strength when
the diffuse double layer is thin compared to the Brinkman screening
length, \ie, when $\kappa \bsl \gg 1$. Under these conditions, the
electrical force (inside the diffuse double layer) is balanced by
viscous stress and, hence, the convective flows are similar. When
$\kappa \bsl = \ka (\bsl / a) = 1$, for example, $\ka \approx 100 /
3.11 \approx 32$ and, as expected, the limiting behavior ($\kappa \bsl
\gg 1$) at high ionic strength occurs when $\ka$ exceeds this value.

\section{Summary} \label{sec:summary}

A rigorous theoretical methodology was presented to calculate steady
electrokinetic transport of electrolytes in a continuous polymer gel
embedded with charged spherical inclusions. Composites with this
microstructure are candidates for enhanced gel-electrophoresis,
chemical sensing, membrane separation, and, perhaps, electroosmotic
pumping technologies. This work was also motivated by a desire to
interpret experiments that probe the surface charge of immobilized
colloids and the micro-rheology of delicate polymer gels.

From a numerically exact treatment of electromigration, diffusion, and
and convective transport past a single inclusion in an unbounded
polymer gel, averaged equations describing bulk transport properties
were derived. The theory was applied to calculate the response to a
steady electric field with a uniform bulk electrolyte
concentration. Note that the response to a bulk electrolyte
concentration gradient will be treated in a forthcoming
publication. In this work, electromigration and diffusion were found
to be independent of convection and, hence, of the polymer-gel (Darcy)
permeability. However, the strength of electroosmotic flow reflects
the gel permeability and, to a lesser (but still significant) extent,
polarization and relaxation by electromigration and diffusion.

When particles are immobilized in a (neutral) polymer gel, the
electrical conductivity is independent of the polymer, whereas the
pore mobility---similarly to the electrophoretic mobility of
dispersions---reflects the size and charge of the inclusions {\em and}
the Darcy permeability. Furthermore, when the Debye length is smaller
(greater) than the Brinkman screening length $\bsl$ (square root of
the Darcy permeability), the pore mobility increases linearly
(quadratically) with $\bsl$.

The variation of pore mobility with $\zeta$-potential and ionic
strength is more complicated because of the significant influence of
polarization and relaxation. Nevertheless, the mobility is
(approximately) inversely proportional to the inclusion radius,
indicating, as expected, that the average flow is proportional to the
average counter-charge density.

The Darcy drag of the intervening polymer gel leads to slow flows that
will often be independent of the differential pressure required to
pump fluid through a (modest) microfluidic network. Optimal pumping
efficiency is favored by thin membranes with large cross-section, high
inclusion volume fractions, and low electrolyte conductivities.

The present model assumes that the polymer gel does not influence ion
mobilities (diffusion coefficients), and that the volume fraction of
the inclusions is small. By analogy with Maxwell's well-known theory
for conduction in dilute random arrays of spheres, the theory advanced
in this work may be accurate at moderate volume fractions. Note also
that the calculations require the bulk electrostatic potential and
electrolyte ion concentrations to vary slowly in space (and
time). Future development of the model will accommodate harmonic
temporal fluctuations in the applied electric field, permitting the
interpretation of dielectric relaxation spectroscopy experiments.

\begin{acknowledgments}
  Supported by the Natural Sciences and Engineering Research Council
  of Canada (NSERC), through grant number 204542, and the Canada
  Research Chairs program (Tier II). The author thanks I. Ispolatov
  (University of Santiago) for fruitful discussions related to this
  work, and an anonymous referee for helpful suggestions.
\end{acknowledgments}

\bibliography{/home/rhill/latex/bibliographies/global}
\bibliographystyle{jfm.bst}

\appendix

\section{Derivation of the averaged momentum equation} \label{app:1}

This appendix supplements~\S\ref{sec:momentum} where the
single-particle microscale problem is adopted to quantify the
influence of the inclusions on bulk momentum transport. Note that the
closures in this work neglect hydrodynamic interactions between
inclusions, which are screened by the intervening Brinkman medium. The
reader is referred to~\cite{Hinch:1977} for details necessary to
account for hydrodynamic interactions. Note also that the fields
obtained from the single-particle problem in this work approximate the
conditionally averaged fields in Hinch's ensemble averaging
methodology. Because the microstructure is homogeneous, the volume
averages below are equivalent to ensemble averages. Integrals over the
surface or volume of a (spherical) inclusion centered at $r = 0$ are
identified by a range of integration that involves the radial
coordinate $r$; otherwise integrals refer to ``representative control
volumes''. Unless stated otherwise, the notation and symbols below are
the same as in the main text.

Outside the inclusions, the (inertialess) microscale momentum
conservation equation is
\begin{equation}
  \vect{0} = \dive{\vect{T}} - (\eta / \bsl^2) \vect{u} - \rho
  \grad{\psi} + \rho_f \vect{g},
\end{equation}
where $\vect{T} = - p \vect{\delta} + 2 \eta \vect{e}$ is the
Newtonian stress tensor, $\rho_f$ is the fluid density, and $\rho_f
\vect{g}$ is a uniform body force (\eg, gravity).

Similarly, let the microscale momentum conservation equation inside
the (rigid) inclusions be
\begin{equation}
  \vect{0} = \dive{\vect{T}^s} + \rho_p \vect{g} + \vect{f}^g,
\end{equation}
where $\vect{T}^s$ is the stress tensor, and $\rho_s$ is the
density. Note that $\vect{f}^g$ is a {\em generalized function} to
represent the electrical and mechanical-contact forces acting on the
surfaces of the inclusions: $\vect{f}^g = \vect{0}$ when $r \ne a$,
and
\begin{equation}
  V^{-1} \int_V \vect{f}^g \mbox{d}V = n \langle \vect{f}^g \rangle
\end{equation}
when $n V \gg 1$ and $V^{1/3}$ is small compared to the characteristic
macroscopic length scale.

Averaging the momentum conservation equation yields Brinkman's
equation for the continuous phase (fluid-saturated polymer gel), with
additional terms arising from the inclusions:
\begin{eqnarray} \label{eqn:1}
  \vect{0} &=& \dive{\langle \vect{T} \rangle} - (\eta / \bsl^2)
  \langle \vect{u} \rangle - \langle \rho \grad{\psi} \rangle + \rho_f
  \vect{g} \nonumber \\ &+& \dive \{ n \langle \int_{r < a}
  (\vect{T}^s + p \vect{\delta}) \mbox{d}V \rangle \} + n \langle
  \vect{f}^g \rangle + \phi (\rho_p - \rho_f) \vect{g}.
\end{eqnarray}

Note that $\langle \rho \grad{\psi} \rangle$ includes only the counter
charge; the influence of the fixed surface charge is captured by
$\langle \vect{f}^g \rangle$, which is the sum of the average
electrical $\langle \vect{f}^e \rangle$ and mechanical-contact
$\langle \vect{f}^m \rangle$ forces acting on the surfaces of the
inclusions (or more generally inside).  Accordingly, the (averaged)
equation of motion for the (immobilized) inclusions is
\begin{equation} \label{eqn:2}
  n \langle \vect{f}^d \rangle + n \langle \vect{f}^g \rangle + \phi
  \rho_p \vect{g} = \vect{0},
\end{equation}
where
\begin{equation} \label{eqn:3}
  \langle \vect{f}^d \rangle = \langle \int_{r=a} \vect{T} \cdot
  \vect{e}_r \mbox{d}A \rangle
\end{equation}
is the average (drag) force exerted by the fluid on the inclusions.

A useful identity for transforming volume to surface integrals is
obtained by differentiating the product (summation on repeated
indices)
\begin{equation}
  \partial (\alpha_{klm...} x_i) / \partial x_j = x_i \partial
  \alpha_{klm...} / \partial x_j + \alpha_{klm...} \delta_{ij},
\end{equation}
where $\alpha_{klm...}$ are the components of an arbitrary-order
tensor (\eg, stress). Integrating this over a volume $\int \mbox{d}V$
enclosed by a surface $\int \mbox{d} A$ and applying Gauss's integral
theorem gives
\begin{equation}
  \int \alpha_{klm...} \mbox{d}V = \int x_i \alpha_{klm...}  \hat{n}_j
  \mbox{d}A - \int x_i \partial \alpha_{klm...} / \partial x_j
  \mbox{d}V.
\end{equation}

Therefore, with $\vect{T}^s = \vect{T}$ at $r = a$, and
$\dive{\vect{T}^s} = - \rho_p \vect{g}$ and $\vect{e} = \vect{0}$
(rigid particles) when $r < a$, it follows that
\begin{equation}  \label{eqn:4}
  \langle \int_{r < a} (\vect{T}^s + p \vect{\delta}) \mbox{d}V
  \rangle = \langle \int_{r=a} \vect{r} 2 \eta \vect{e} \cdot
  \vect{e}_r \mbox{d}A \rangle.
\end{equation}

In general, body forces and the isotropic stress contribute $\langle
\int \vect{r} (\vect{f}^g + \rho_p \vect{g} - \grad{p}) \mbox{d}V
\rangle$ to the right-hand side of Eqn.~(\ref{eqn:4}), but these
integrals vanish if, on average, the (internal) body-force
distributions and (internal) pressure gradient are even functions of
position. With a macroscale velocity gradient (not considered in this
work), the extensional and rotational contributions to $\vect{e}$ are
even functions of position, so $\langle \int_{r=a} \vect{r} \vect{e}
\cdot \vect{e}_r \mbox{d}A\rangle$ will be linear in $\langle
\grad{\vect{u} } \rangle$ and, therefore, will contribute to the bulk
(deviatoric) stress by modifying the effective viscosity $\eta'$ (\eg,
the well-known Einstein viscosity for dilute, force-free suspensions).

Finally, collecting the results from
Eqns.~(\ref{eqn:1})--(\ref{eqn:4}) gives
\begin{eqnarray} \label{eqn:5}
  \vect{0} &=& \dive{\langle \vect{T} \rangle}- (\eta / \bsl^2)
  \langle \vect{u} \rangle - \langle \rho \grad{\psi} \rangle + (1 -
  \phi) \rho_f \vect{g} \nonumber \\ &+& \dive{ \{ n 2 \eta \langle
  \int_{r=a} \vect{r} \vect{e} \cdot \vect{e}_r \mbox{d}A \rangle \} }
  - n \langle\vect{f}^d \rangle.
\end{eqnarray}

The correctness of Eqn.~(\ref{eqn:5}) can be verified, in part, by
considering a stationary fixed bed of inclusions in the absence of
electrical forces. Accordingly, Eqn.~(\ref{eqn:5}) simplifies to
\begin{equation} \label{eqn:7}
  \vect{0} = - \grad{\langle p \rangle} + (1 - \phi) \rho_f \vect{g}
  \nonumber - n \langle\vect{f}^d \rangle,
\end{equation}
and the static equillibrium of the inclusions requires
\begin{equation}
  \vect{0} = n \langle \vect{f}^d \rangle + n \langle \vect{f}^g
  \rangle + \phi \rho_p \vect{g}.
\end{equation}
Therefore,
\begin{equation}
  \vect{0} = - \grad{\langle p \rangle} + n \langle \vect{f}^g \rangle
  + \rho_f \vect{g} + \phi(\rho_p - \rho_f) \vect{g},
\end{equation}
and, hence, in a stationary fluid where $\grad{\langle p \rangle} =
\rho_f \vect{g}$, the average force required to immobilize the
inclusions is simply
\begin{equation}
  \langle \vect{f}^g \rangle = (\phi / n) (\rho_f - \rho_p) \vect{g} =
  (4/3) \pi a^3 (\rho_f - \rho_p) \vect{g}.
\end{equation}

\end{document}